\documentclass[
 reprint,
 amsmath,amssymb,
 aps,
 superscriptaddress
]{revtex4-2}
\usepackage{xcolor}
\usepackage{graphicx}
\usepackage{dcolumn}
\usepackage{bm}
\usepackage{hyperref}
\usepackage[mathlines]{lineno}

\DeclareMathOperator{\diag}{diag}
\DeclareMathOperator{\Tr}{Tr}

\DeclareMathOperator{\erf}{erf}
\DeclareMathOperator{\E}{\mathbb{E}}
\usepackage{slashed} 
\usepackage{bm} 
\usepackage{breqn}
\usepackage{mathtools}
\usepackage{multirow} 
\usepackage{indentfirst} 
\usepackage{lipsum} 
\usepackage{mathrsfs} 
\usepackage{float} 
\usepackage{subfigure}
\DeclareMathAlphabet{\mathbbb}{U}{bbold}{m}{n} 
\usepackage{csquotes} 
\usepackage{enumitem} 
\usepackage{cuted}

\usepackage{comment} 
\newcommand{\nn}{\nonumber}
\begin{document}

\preprint{APS/123-QED}


\title{Baryon and meson masses in the Nambu--Jona-Lasinio model: a Bayesian approach}

\author{Antoine Pfaff}
\email{a.pfaff@ip2i.in2p3.fr}
\affiliation{Univ. Lyon\char`,{} Univ. Claude Bernard Lyon 1\char`,{}
CNRS/IN2P3\char`,{} Institut de Physique des 2 Infinis de Lyon\char`,{} UMR 5822\char`,{} 69622 Villeurbanne\char`,{} France
} 
    
\affiliation{SUBATECH\char`,{} Nantes University\char`,{}  IMT Atlantique\char`,{}  IN2P3/CNRS\char`,{}  \\ 4 rue Alfred Kastler\char`,{}  44307 Nantes cedex 3\char`,{}  France
}

\author{Hubert Hansen}
\affiliation{Univ. Lyon\char`,{} Univ. Claude Bernard Lyon 1\char`,{}
CNRS/IN2P3\char`,{} Institut de Physique des 2 Infinis de Lyon\char`,{} UMR 5822\char`,{} 69622 Villeurbanne\char`,{} France
}

\author{Joerg Aichelin}
\affiliation{SUBATECH\char`,{} Nantes University\char`,{}  IMT Atlantique\char`,{}  IN2P3/CNRS\char`,{}  \\ 4 rue Alfred Kastler\char`,{}  44307 Nantes cedex 3\char`,{}  France
}

\author{Juan M. Torres-Rincon}
\affiliation{Departament de F\'isica Qu\`antica i Astrof\'isica and Institut de Ci\`encies del Cosmos (ICCUB)\char`,{}  Facultat de F\'isica\char`,{}   Universitat de Barcelona\char`,{}  Mart\'i i Franqu\`es 1\char`,{}  08028 Barcelona\char`,{}  Spain}
    
\date{\today}

\begin{abstract}

We investigate the capabilities of the Nambu--Jona-Lasinio model to describe and reproduce fundamental vacuum properties of Quantum Chromodynamics, notably the hadronic spectrum. Mesons are described as quark-antiquark bound states at the level of the random phase approximation of the Bethe-Salpeter equation, while baryons are characterized as quark-diquark bound states within the static approximation of the Faddeev equation.
Within a Bayesian framework, we constrain the model by phenomenologically known quantities and study the implications on its parameters and predictions in vacuum, as well as the correlations between the two. We find that within our framework, the vacuum masses of mesons and baryons can be reasonably well reproduced. Scalar diquarks need to be significantly bound in order to correctly reproduce the masses of the baryon octet, therefore enforcing values of the scalar diquark coupling larger than what is suggested by the canonical Fierz values. These findings could have important implications on the phenomenology of strongly-interacting matter
at high temperature and density as well as of compact star physics.
\end{abstract}
\maketitle

\section{\label{Introduction}Introduction}

Modeling of hadrons---and their possible composite nature---is a challenging task because their internal structure lies out of the realm of perturbative Quantum Chromodynamics (QCD). In addition, they are confined color neutral objects and confinement is one of the least understood properties of QCD. Hadrons are, however, in the center of interest of current research on physical systems composed by strongly-interacting particles. Besides, these systems can be exposed to a finite temperature---like the medium formed in relativistic heavy-ion collisions---or to finite density---such as the nuclear medium, or the interior of neutron stars.  The study of hadrons in matter is therefore in the center of interest of present nuclear physics.

An attractive possibility is to solve the QCD field equations on a discretized lattice using a fully numerical approach. This idea has been very successful to extract hadron properties, both in vacuum~\cite{Durr:2008zz,Fodor:2012gf,Dudek:2013yja,Detmold:2019ghl} and at finite temperature~\cite{Aarts:2017rrl,Bazavov:2019www,Garcia-Mascaraque:2021sbt}. The latest lattice-QCD (LQCD) calculations for the light sector can in most cases approach the ``physical point'' (with a physical pion mass ${m_\pi \approx 138}~$MeV), and the extrapolation to the continuum limit can address the spectroscopy in a rather faithful way~\cite{RBC:2014ntl,Fischer:2020yvw,Paul:2021pjz}. However, calculations of the baryon spectrum reaching the physical point have not been implemented yet.
In addition, LQCD calculations are computationally costly, and cannot be straightforwardly extended to finite baryon densities due to the so-called ``numerical sign problem''.

Another option to address hadron spectroscopy is the application of (truncated) QCD Dyson-Schwinger and Bethe-Salpeter equations which can accommodate both finite temperature and density (see Ref.~\cite{Roberts:2000aa} and references therein). Results of this approach on vacuum spectroscopy are presented in the review~\cite{Eichmann2016}, while at finite temperature some results for light mesons have been recently  published~\cite{Dorkin:2018aap,Gao:2020hwo}.

In this work we take the approach of QCD-inspired models. In particular, we would like to describe a substantial set of hadrons---including mesons and baryons---but incorporating the minimal set of degrees of freedom and parameters in the model. We also would eventually like to extend the calculations to finite temperature and density (while in this work we restrict ourselves to the vacuum case).
With this goal in mind, we focus on the Nambu--Jona-Lasinio (NJL) model: a low-energy approach of QCD which describes interactions among quarks while gluons are integrated out. The final interactions result (in the simplest local version of the model) in multiquark contact terms with an energy-independent strength. This interaction can be tuned to reproduce the physical value of several observables in the low-energy hadron sector.

The NJL model has been profusely used in the literature---both in vacuum and in a thermal/dense medium---due to its technical simplicity and its properties \cite{Klevansky,Hatsuda,Klimtreview,Vogl1991,Rehberg_su3,Buballa}: it describes the spontaneous symmetry breaking property of QCD and the generation of (pseudo) Goldstone bosons, it can easily be extended to include in-medium properties, it illustrates the chiral symmetry restoration at finite temperatures/densities, it can accommodate the effects of the U(1)$_A$ axial anomaly at the meson level ($\eta-\eta'$ mass splitting) and, using the Bethe-Salpeter or the Faddeev equations, it can describe the dynamical generation of colorless mesons and baryons, which can be identified with physical hadrons.

With respect to the last point, the local NJL model is able to generate the masses of ground state mesons (scalar, pseudoscalar, vector and axial vectors) as bound states in the quark-antiquark scattering, as well as baryons (members of the SU$(3)_f$ octet and decuplet) as bound states of a quark and a diquark (the latter itself being a bound state of two quarks). The original formalism is based on the relativistic Faddeev equation for three quarks which, under certain approximations, can be simplified to a two-body equation~\cite{Reinhardt:1989rw,Buck:1992wz}. Using the NJL model, some vacuum baryon masses were already obtained in~\cite{Buck:1992wz,Ebert:1994mf,Ishii:1995bu}, and more recently, as functions of temperature and/or density in~\cite{Gastineau:2002jh,Wang:2010iu,Blanquier2011,Mu:2012zz,TorresRincon2015}---notice that at finite temperature/density one can also use the Polyakov loop-extended NJL (PNJL) model~\cite{2004,Ratti2006}, which incorporates certain gluonic components, allowing to suppress part of the unphysical quark modes in the partition function at low $T$ and $n$.

In a previous publication~\cite{TorresRincon2015}, some of us studied the thermal evolution of several baryons with three flavors with the NJL and PNJL models. We were able to satisfactorily reproduce the vacuum masses of 8 baryon states ($N,\Delta,\Sigma,\Lambda,\Xi,\Sigma^\star,\Xi^\star,\Omega$) with only two additional parameters in the NJL model (scalar and axial-vector couplings within the diquark sector). This was achieved without spoiling the good agreement of meson masses, but at the cost of having a sizable light quark dressed mass of $m_q \simeq 480$ MeV. Being not an observable, such a value for the quark mass is not a particularly troubling result---a typical quark mass would be of the order of $m_\Delta/3 \simeq 400$ MeV, as it seems to be required in previous calculations~\cite{Buck:1992wz,Ebert:1994mf,Ishii:1995bu}---but it was unclear whether this was a necessary condition of the model in order to fit the meson and baryon spectroscopy, or whether a different parametrization with a smaller quark mass could also be found.

In this paper we employ the Bayesian formalism to perform a systematic analysis of the NJL model prediction for the vacuum hadron spectrum. The Bayesian approach has notably been implemented with the NJL model in the context of the study of the compact star equation of state, where the models are conditioned by measurements of the star masses, radii and tidal deformabilities~\cite{Alvarez2016,Blaschke2021,Shahrbaf:2021cjz,pfaff1}.
However, in the current work the parametrization of the model is instead constrained by a likelihood informed with the experimentally measured masses of the full pseudoscalar meson octet, baryon octet and baryon decuplet as well as the pion and kaon weak decay constants. 
Our goals in this work are: (i) Find whether the minimal version of the NJL model (local, mean field and with the minimal number of couplings) is able to provide a good matching of these quantities overall in a systematic approach, (ii) obtain and analyze the actual correlations between parameters, and between parameters and observables, (iii) address the question whether a good fit of the hadron spectrum in the NJL model requires to have a relatively massive quark, as indicated in the recent studies~\cite{Blanquier2011,Blanquier2013,TorresRincon2015}, and (iv) sketch an ``optimal'' parametrization which can be used in future studies of hadron spectrum and the phase diagram of the model at finite temperature and density.

This paper is organized as follows: In Sec.~\ref{sec::model} we introduce our minimal version of the NJL model and detail the level of approximations employed to compute both meson and baryon vacuum masses. 
In Sec.~\ref{sec::res} we describe the Bayesian method that we apply to explore the full parameter space and confront the outcome of the model to the experimental results. Finally, in Sec.~\ref{sec::discussion} we analyze the posterior distributions of the parameters and physical observables (with a particular emphasis on the diquark and baryon sector), and comment the most interesting correlations found. We present a summary of this work in Sec.~\ref{sec::summary}, and conclude the paper with a set of appendices to which all the technical details are referred.

\section{Nambu--Jona-Lasinio Model}\label{sec::model}

Our analysis is based on the description of the low-energy hadronic spectrum of the three-flavor Nambu--Jona-Lasinio (NJL) quark model \cite{Klevansky,Hatsuda,Klimtreview,Vogl1991,Rehberg_su3,Buballa}, with the following Lagrangian density,

\begin{align}
\mathcal{L}_{ \textrm{NJL} } & = \overline{q}(i \slashed{\partial}-\hat{m}_0) q + G \ [(\overline{q}\tau^{a}q)^2 + (\overline{q}i\gamma_5\tau^{a}q)^2] \nn \\
& -K\Big(\operatorname{det}\limits_{f} [\overline{q}(1+\gamma_5)q]+\operatorname{det}\limits_{f} [\overline{q}(1-\gamma_5)q] \Big) \nn \\
&+G_{d} \ (\overline{q}i\gamma_5C\tau^{A}\lambda^{A'}\overline{q}^{T})(q^{T}Ci\gamma_5\tau^{A}\lambda^{A'}q) \nn \\
&-G_{d,V} \ (\overline{q}\gamma^\mu C\tau^{S}\lambda^{A'}\overline{q}^{T})(q^{T}C\gamma_\mu\tau^{S}\lambda^{A'}q) \ . \label{lagrangian}
\end{align} 

In Eq.~(\ref{lagrangian}) we denote $\tau^a$, ${a=0,1,...,8}$ the $\mathfrak{su}(3)$ Gell-Mann matrices in flavor space (with ${\tau^0=\sqrt{\frac{2}{3}}\mathbbb{1}_f}$) and as $\lambda^{a'}$, ${a'=0,1,...,8}$ the $\mathfrak{su}(3)$ matrices in color space (with $\lambda^0=\sqrt{\frac{2}{3}}\mathbbb{1}_c$). $A$ and $A'$ are summed over the indices of antisymmetric Gell-Mann matrices; $A,A^'=2,5,7$; while $S$ is summed over the indices of symmetric Gell-Mann matrices; ${S=0,1,3,4,6,8}$. In the diquark sector---last two lines of Eq.~(\ref{lagrangian})---the charge conjugation matrix is given by $C=i\gamma_0\gamma_2$.

The quarks are described by the flavor triplet spinor field ${q=(u,d,s)}$ with bare masses ${\hat{m}_0 = \diag(m_{0,u}, m_{0,d},m_{0,s})}$. Quarks interact~with each other through four- and six-fermion contact vertices controlled by four coupling constants: the scalar-pseudoscalar coupling $G$, the 't\,Hooft coupling $K$, the scalar diquark coupling $G_d$, and the axial diquark coupling $G_{d,V}$. 

The NJL Lagrangian shown in Eq.~(\ref{lagrangian}) obeys the global {SU(3)$_R\times$SU(3)$_L\times$U(1)$_B$} symmetry of QCD in the limit of vanishing bare quark masses ($\hat{m}_0=0$), and substitutes its local color gauge symmetry with a global invariance under SU$(3)_c$ transformations.
The so-called axial anomaly of QCD is accounted for by the presence of the 6-fermion Kobayashi--Maskawa--'t\,Hooft interaction term, which breaks the U(1)$_A$ symmetry explicitly \cite{KobayashiMaskawa,thooft1,thooft2}.
We neglect the quark-antiquark vector channel as it does not have any effect in the vacuum at the mean-field level, and we checked that the vector meson spectrum is vastly uncorrelated to the rest of the properties of the model, yielding only minor physical constraints. 
Although other diquark channels are in principle needed in order to meet the symmetry requirements, they are not considered in this work as they do not contribute to the baryons in our scheme.

The form of the four-fermion part of the NJL Lagrangian can be directly derived by carrying out a Fierz transformation of a global color current-current interaction, which can be seen as an approximation of QCD in the limit of low gluon momentum exchange~\cite{Buballa}.
Following this procedure, four-fermion interaction terms in all possible spin, flavor and color channels emerge, and we only kept in Eq.~(\ref{lagrangian}) the ones that give rise to the low-mass hadronic spectrum of interest. 
Since the original theory of QCD has only one coupling constant (the strong coupling $g_s$), the Fierz method yields
fundamental relationships between the coupling constants in each interacting channel~\cite{TorresRincon2015,Klevansky,Buballa}. For the terms appearing in Eq.~(\ref{lagrangian}) and for $N_c=3$, the relevant Fierz relations are $G_d = 3G/4 $ and $G_{d,V}= 3G/8 $~\cite{Buballa}. In this work however, we will forego these relations and keep $G_d$ and $G_{d,V}$ as free parameters. As we shall see below, the above relations are not compatible with the experimental masses of the baryon spectrum in our framework.

From the Lagrangian of Eq.~(\ref{lagrangian}), the dressed mass $m_f$ of the quark of flavor $f$ can be obtained by solving the self-consistent Dyson equation \cite{Dyson1949} for the quark propagator, which in the Hartree approximation eventually yields:
\begin{equation} \label{gap}
    m_i = m_{i,0}-4G\langle\overline{q}_iq_i\rangle +2K\langle\overline{q}_jq_j\rangle\langle\overline{q}_kq_k\rangle \ ,
\end{equation}
where ($i$,$j$,$k$) denotes any permutation of ($u$,$d$,$s$).
The chiral condensates $\langle\overline{q}_fq_f\rangle$ are related themselves to the quark masses by the single line loop integral:
\begin{equation}\label{chircond}
    \langle\overline{q}_fq_f\rangle =  - i \int \frac{d^4k}{(2\pi)^4}\Tr S_f(k) \ ,
\end{equation}
where the traces are performed in color and spinor spaces, and $S_f(p)$ is the dressed quark propagator, given at zero temperature by

\begin{equation} \label{quarkpropagator}
    S_f(k) = \frac{\slashed{k}+m_f}{k^2-m_f^2+i\varepsilon}
\end{equation}
After performing the traces, one simply obtains
\begin{equation}\label{chircond2}
\begin{split}
    \langle\overline{q}_fq_f\rangle 
    & = -4i m_f N_c \int^\Lambda\frac{d^4k}{(2\pi)^4}\frac{1}{k^2-m_f^2+i\varepsilon}\\
    &= -4i m_f N_c I_1(m_f,\Lambda) 
    \end{split} \ ,
\end{equation}
Since the integral $I_1$ is divergent, it must be regularized by the introduction of an ultraviolet cutoff $\Lambda$, whose scale is related to the asymptotic freedom scale of QCD. In this work, we use a simple 3-momentum noncovariant regularization scheme in which the cutoff $\Lambda$ is imposed after carrying out the $k_0$ integration. Details on the evaluation of the integrals are relegated to the App.~\ref{app:I1}.

Mesons and diquarks can be described within the NJL model by solving the Bethe-Salpeter equation for the scattering of a quark and an antiquark (or two quarks in the case of diquarks). The existence of a bound state can then be inferred from the $T$-matrix.
In the so-called ring approximation (random phase approximation) for the $T$-matrix, the infinite resummation of diagrams is reduced to a geometric sum involving the quartic coupling and the two-particle loop function~$\Pi_X$(p),
\begin{align}
T_{X}(p) &= \mathcal{G}_X + \mathcal{G}_X\Pi_X(p)\mathcal{G}_X + ... \nn \\
&= \frac{\mathcal{G}_X}{1-\mathcal{G}_X\Pi_X(p)} \ , \label{geomsum}
\end{align}
where $\mathcal{G}_X=2G_X$ is, up to a loop factor 2, the four-quark coupling constant appearing in the Lagrangian, that selects the appropriate interaction channel associated with the quantum numbers $X$. $T_{X}$ refers to the $T$-matrix in the channel $X$, which can describe a bound state if a pole in the physical Riemann sheet of $p$ is generated. 
The mass of the bound state can be inferred as the pole position in the real energy axis. Therefore, we identify the mass $m_X$ (at rest) of the meson/diquark $X$ as the solution of the equation,
\begin{equation}
1-\mathcal{G}_X\Pi_X(p)|_{p_0=m_X , \,\bm{p}=0}=0 \ .
\end{equation}
This pole can in principle lie outside of the real axis, reflecting a finite width related to a decay into a quark-(anti)quark pair. If this is the case, we will identify the mass as the real part of the pole.
One should keep in mind, however, that such a decay is unphysical and only reflects the lack of confinement of the NJL model and its description of bound states. In addition, we will consider stable bound states in almost all cases, such that their masses are indeed real. 

Each bound state $X$ can be associated with an appropriate quark-antiquark (or quark-quark in the case of diquarks) interaction channel $\Gamma_X$ depending on the quantum numbers exchanged through the interaction vertex (\textit{e.g.} the pseudoscalar pion $\pi^0$ is associated with the vertex ${\Gamma_{\pi^0}  = i\gamma_5 \tau_3\mathbbb{1}_c}$).
The expression of the loop function $\Pi_X(p)$ can be obtained using the Feynman rules for fermions, and is given by
\begin{equation}
\Pi_X(p)=i\int^\Lambda\frac{d^4k}{(2\pi)^4}\Tr\Big(\Gamma_X S(p+k) \Gamma_X^\dagger S(k)\Big) \ ,
\end{equation}
where ${S(k) = \diag(S_u(k),S_d(k),S_s(k))}$ and the trace must be performed in color, flavor and spinor spaces.
Further details on the calculation of this loop function in the meson and diquark channels of interest are given in App.~\ref{Appendixpol}. 

For each bound state, one can additionally compute its coupling strength $g_{X\rightarrow qq'}$ corresponding to the residue of the propagator of bound state $X$ at its pole
\begin{equation}
g_{X\rightarrow qq'}^2= \Big(\frac{\partial\Pi_X}{\partial p^2}\Big|_{p^2=m_X^2}\Big)^{-1} \ .
\end{equation}
Around its pole, the $T$-matrix element of the bound state~$X$ can then be expressed in the form
\begin{equation}\label{propagapprox}
   T_X(p) \approx -\frac{g_{X\rightarrow qq'}^2}{p^2-m_X^2} \ ,
\end{equation}
which can readily be interpreted as the bound state propagator.
Note that the equations above apply strictly speaking only for spin 0 bound states. In the case of spin~1 (in particular for the axial diquarks), a similar procedure can be applied in order to take into account the Lorentz structure of the propagator~\cite{TorresRincon2015,Klimtreview}.

Finally, the axial decay constants $f_{X}$ can be obtained from the evaluation of the transition matrix element between a meson and the vacuum.
\begin{equation}
   i f_{X}p^\mu= \langle 0 | j^{\mu}_{5,X} | X \rangle \ .
\end{equation}
In the NJL model, the evaluation of the diagram at one-loop level gives~\cite{Klevansky}
\begin{equation} \label{decayconst}
    i f_{X}p^\mu = i\frac{g_{X\rightarrow qq'}}{2}\int^\Lambda\frac{d^4k}{(2\pi)^4}\Tr\Big(\gamma^\mu\Gamma_XS(p+k)\Gamma_X^\dagger S(k)\Big) \ ,
\end{equation}
which is computed as explained in App.~\ref{app:fX}. In this work, we will only focus on the pion and kaon decay constants which are experimentally well-known quantities.

In an analogous way to the modelling of mesons as quark-antiquark bound states, baryons can be constructed by solving the so-called Faddeev equation~\cite{Faddeev:1960su} for the scattering of three quarks.  While even in the NJL model this equation is difficult to solve, the quark-diquark framework provides an appealing approximation in order to simplify the problem (see~\cite{Eichmann2016} and references therein). In this case, the three body scattering problem can be simplified to a two body problem with a quark and a diquark, in which case the Faddeev equation reduces to a Bethe-Salpeter like equation, and the interaction between a quark and a diquark is brought down to a simple quark exchange. 

To simplify further the calculation, we apply the ``static approximation'' consisting in neglecting the momentum carried by the exchanged quarks in the diagram summation~\cite{Buck:1992wz}. The calculation is then reduced to a geometric sum very similar to Eq.~(\ref{geomsum}). 
Such an approximation is valid as long as the in-medium mass of the quark remains high and the chiral symmetry remains spontaneously broken. At high temperature and density, chiral symmetry should be restored and the static approximation is not justified anymore. 
However, since in this work we are only interested in studying chiral properties in the vacuum, this approximation should remain valid. We refer to Ref.~\cite{TorresRincon2015} and App.~\ref{appbaryon} for further details on the calculation.

Since diquarks are colored objects, each baryon wavefunction must be decomposed with a specific superposition of quark and diquark combinations in order to ensure that the constructed baryons are colorless. Baryons of the octet (with $J^\pi=\frac{1}{2}^+$) are assumed to be constituted only of scalar diquarks, while baryons of the decuplet (with $J^\pi=\frac{3}{2}^+$) are assumed to be constituted of axial diquarks. The corresponding projection for each baryon can be found in Ref.~\cite{TorresRincon2015,Hanhart_1995}, following the principle that the total baryon wave function should be antisymmetric in its decomposition.

\section{Methodology} \label{sec::res}

The purpose of this work is to investigate the parameter dependence of the NJL model predictions, as well as to understand how to choose these parameters in order to be in agreement with experimentally measured properties of the hadronic spectrum. To do so, we adopt a Bayesian methodology in order to constrain Monte Carlo generated models to satisfy the physical constraints. A large number of NJL parameter sets is generated within a large and uninformed (uniform) prior. The predictions of each set are then computed and confronted with experimental data (here observables of the SU$(3)_f$ hadronic spectrum) to select parameter sets that are consistent with real world physics. All calculations are performed in vacuum.

Our framework based on the NJL model comprises a total of seven different parameters; two bare masses for the light and strange quarks (we work in the isospin limit  where $m_0 \equiv m_{0,u}=m_{0,d}$), four coupling constants ($G$, $K$, $G_d$ and $G_{d,V}$), and one momentum cutoff $\Lambda$. In other works using a similar framework \cite{Gastineau2002,Osipov:2004bj,Mu:2012zz,Ebert2005,Lawley_2006,TorresRincon2015,Blanquier2011}, these parameters were partially or totally fine tuned in order to reproduce several phenomenological quantities of the hadronic spectrum. However, this method possesses some drawbacks:
\begin{enumerate}[label=(\roman*)]
    \item there are more than seven quantities that can be fitted from hadron phenomenology, meaning that the choice of the fitted quantities can affect the results significantly and that the resulting parameter set is not necessarily optimal for reproducing known experimental quantities,
    \item this method does not reflect the uncertainties (both experimental and theoretical) that can blur the model inputs.
    \item this method cannot account for the correlations between the model parameters and the observables, which carry significant information on the relationships between physical sectors of the model.

\end{enumerate}

For these reasons, we propose a method to consistently take into account the uncertainties on all available quantities.

\subsection{Bayesian statistical analysis}

The core of a Bayesian statistical analysis lies on Bayes' theorem in probability theory~\cite{Sivia2006}. Assuming a set of observables $\{o_i\}$ constrained by some data, the theorem states

\begin{equation}
    \mathcal{P}(\{o_i\}| {\rm data}) = \mathcal{N} \ \mathcal{P}( {\rm data}|\{o_i\})\ \mathcal{P}(\{o_i\}) \ .
\end{equation}
It relates the \textit{posterior} probability of a model $\mathcal{P}(\{o_i\}| {\rm data})$ to both its \textit{prior} probability $\mathcal{P}(\{o_i\})$, and to the \textit{likelihood} function $\mathcal{P}( {\rm data}|\{o_i\})$ . 
The quantity $\mathcal{N}$ is a normalization factor also known as \textit{evidence}. Here, each individual parameter set (which will be also denote as ``individual model'', as it is the accepted term in this framework), 
is represented by the collection of predictions it makes on the observables $\{o_i\}$. 

The prior probability represents the initial knowledge that we have on the observables $\{o_i\}$. 
It is entirely determined by the modeling of the observables and the amount of freedom that is allowed within this modeling. 
In order to perform a study that is the most agnostic possible and span the largest uncertainties, the prior distribution should remain as uninformative as possible. 

The likelihood estimates the compatibility of the model with the data by attributing a probability to the experimental results (the data) for each possible underlying model. In this sense, it acts as a filter on the different models considered, by performing a selection to give a preference to the models that reproduce best the data. Its effects can be dependent on the degree of trust one can put on the available experimental results relevant to the analysis.

The posterior probability distribution illustrates the degree of credibility attributed to a model given the quality of its reproduction of experimental data. This distribution can be marginalized over a subset of observables to obtain marginal probability distribution functions (PDFs) on a single quantity~$o_j$

\begin{equation}
    \mathcal{P}(o_j| {\rm data}) = \bigg(\prod_{i\neq j} \int do_i \bigg) \ \mathcal{P}(\{o_i\}| {\rm data}) \ .
\end{equation}

\subsection{Numerical setup}

\subsubsection{Prior knowledge}

The first step consists in generating an uninformative prior on the parameters. We span the model uncertainty by drawing each of the seven model parameters random and independently, using a uniform prior distribution. 
The interval boundaries considered for each of the model parameters are summarized in Table~\ref{NJLprior}. 
These ranges were chosen heuristically from previous SU$(3)_f$ parametrizations that have been suggested in the literature using similar models \cite{Rehberg_su3,TorresRincon2015,Blanquier2013}. Values far outside of these boundaries may typically lead to unphysical behavior (unbound mesons or baryons, absence of resonances, unusually large chiral condensate...). One can check a posteriori that they indeed span a satisfactory uncertainty on the hadron spectrum masses.
Note that we choose as independent variables the quantities {($G\Lambda^2$,~$G_d/G$,~$G_{d,V}/G$)} instead of {($G$,~$G_d$,~$G_{d,V}$)} for the convenience of working with dimensionless quantities that are easier to interpret. 

\begin{table}[htbp]
\centering

\begin{tabular}{|c|c|c|c|c|c|c|c|}

\hline
Parameter & $\Lambda$ & $G\Lambda^2$& $K\Lambda^5$ & $m_0$ & $m_{0,s}$ &$G_d/G$ &$G_{d,V}/G$\\
  &  (MeV) & & & (MeV)& (MeV) & &   \\
\hline
Min& 550 & 1.75 &4 & 4 & 95 &1.0 &0.75 \\
Max & 625 & 4 & 15 &7 &150 &1.89 &1.3 \\
\hline

\end{tabular}
\caption{Minimum and maximum values employed for the prior on each of the independent parameters of the SU$(3)_f$ NJL model.}
\label{NJLprior}
\end{table}

From the prior distribution, we first generate a large number ($N=5\times10^7$) of random parameter sets. For each one of these parameter sets, the self-consistent Dyson equations are solved in order to find the vacuum quark masses, and the masses of the meson pseudoscalar octet as well as baryons from the octet and decuplet are computed by finding the poles of the corresponding propagator using the methods described above. All the models that cannot account for the full set of the chosen hadrons are considered as unrealistic; if any of the hadronic states is either found not to exist or to be unbound (unstable) within the model, the model is discarded from the analysis. This enforces that our prior model population is always able to predict at least the existence of the correct spectrum for the mesons and baryons of interest.
This ``stability condition'' imposes
\begin{equation}\label{stable}
    0 \leq m_X \leq m_{f_1}+m_{f_2} 
\end{equation}
for the mesons and diquarks constituted of a quark and (anti)quark with flavors $f_1$ and $f_2$. For the baryons, the analog condition reads
\begin{equation} \label{baryonstab}
    0\leq m_B \leq m_{d}+m_{q} \ ,
\end{equation}
where $d$ and $q$ represent the diquark and quark that compose the baryon $B$. Both of these conditions ensure that the calculated bound state masses are found on the real axis \footnote{The $\eta^\prime$ meson is oftentimes found to be unstable within this model and violates Eq.~(\ref{stable}), as its mass remains quite large due to the U(1)$_A$ anomaly~\cite{Rehberg_su3}. We will therefore make an exception for this meson and compute its mass as the real part of the complex pole of the $T$-matrix. The associated width is an artifact due to the lack of confinement in the NJL model.}. Note that if a baryon state is constructed as a superposition of several quark-diquark states, we enforce the inequality (\ref{baryonstab}) for all the quark-diquark pairs involved.

\subsubsection{Bayesian likelihood from experimental data}\label{sec::likelihood}

In order to select the models that are the most consistent with phenomenological knowledge, likelihood weights are assigned depending on their quality to reproduce the low-energy hadronic spectrum properties. The quantities that we compute and confront with experimental data are
\begin{enumerate}[label=(\roman*)]
\item the masses of the pseudoscalar mesons: $m_\pi$, $m_K$, $m_\eta$, $m_{\eta^\prime}$,
\item the pion decay constant and the ratio between kaon/pion decay constants: $f_\pi$, $f_K/f_\pi$ ,
\item the light quark chiral condensate: $\Sigma=-\langle\overline{q}q\rangle^{\frac{1}{3}}$ ,
\item the masses of the baryon octet: $m_N$, $m_\Lambda$, $m_\Sigma$, $m_\Xi$, 
\item the masses of the baryon decuplet: $m_\Delta$, $m_{\Sigma^\star}$, $m_{\Xi^\star}$, $m_\Omega$.
\end{enumerate}
Note that since we assume isospin symmetry, isospin degenerate states have been gathered into the same state. In principle, scalar mesons could also be used to further constrain the parameters, but it is known that such resonances are very poorly described in terms of pure quark-antiquark states and without the inclusion of confining effects~\cite{CSWSX95}. For instance, the $f_0(500)$ scalar 
is rather believed to be better characterized by a two-pion state, which our treatment of the NJL model cannot account for at the mean field level due to our lack of description of the two pion continuum.
The scalar masses are in addition more difficult to measure and are therefore not very well constrained by experiments~\cite{Pelez_2016}. 

We determine the likelihood by assigning to the models a Gaussian weight for each quantity $o_i$ (except for the chiral condensate; see below),
\begin{equation}
    w_{i}=\frac{1}{\sqrt{2\pi}\sigma_i}\exp\Big[ -\frac{(o_i-\overline{o_i})^2}{2\sigma_i^2} \Big] \ ,
\end{equation}
where $\overline{o_i}$ is the recommended empirical value for the quantity $o_i$, and $\sigma_i$ the associated uncertainty. 
The total likelihood function is obtained by combining the weights associated with each quantity,
\begin{equation}\label{eq::likelihood}
    \mathcal{P}( {\rm data}|\{o_i\})=\prod_i w_{i} \ ,
\end{equation}
where the index $i$ runs over all physical quantities that need to be constrained. 
All the mean values $\overline{o_i}$ and standard deviations $\sigma_i$ used to implement the Gaussian filters are detailed in Table~\ref{PDGspectrum}. We discuss now how these quantities were determined.

For the pion and kaon masses, these values are not necessarily straightforward to choose since the neutral pion (kaon) and the charged pion (kaon) do not have the same mass. According to the Particle Data Group (PDG), the experimental masses are~\cite{PDG}:
\begin{equation}
        m_{\pi^{0}}=134.9768(5)~\text{MeV} ,\quad   m_{\pi^{\pm}}=139.57039(18)~\text{MeV} \ ,
\end{equation}
\begin{equation}
        m_{K^{0}}=497.611(13)~\text{MeV} ,\quad   m_{K^{\pm}}=493.677(16)~\text{MeV} \ .
\end{equation}
Since the experimental uncertainties are very small, we can neglect them compared to systematic uncertainties arising from the various approximations made within the model. However, since we assume $m_{u,0}=m_{d,0}$ and hence all our results are isospin independent, there is a problem of identification of the NJL pion/kaon with the real life hadrons. In the real world, the masses of $\pi^0$ and $\pi^\pm$ are the result of the combination of strong, electromagnetic and weak effects, while in the NJL description only the strong interaction is taken into account, leading to a degeneracy in our approach. 
This problem is discussed in details in Ref.~\cite{Dmitrasinovic1992,Dmitrasinovic1995} where electromagnetic corrections were applied to the NJL model to describe the mass splitting of the pions. This is however beyond the scope of this work. Therefore, we estimate the mass of the isospin symmetric state to a barycenter of the physical states, and the systematic uncertainty is simply related to the mass splitting of the states:
    \begin{equation}
    \overline{m_\pi}=\frac{2m_{\pi^\pm}+m_{\pi^0}}{3}\,,\qquad \sigma_{m_\pi}=m_{\pi^\pm}-m_{\pi^0}
    \end{equation}
    \begin{equation}
    \overline{m_K}=\frac{m_{K^\pm}+m_{K^0}}{2}\,,\qquad \sigma_{m_K}=m_{K^\pm}-m_{K^0}
    \end{equation}

For the $\eta$ and $\eta^\prime$ mesons, the PDG reports:
\begin{equation}
        m_{\eta}=547.862(17)~\text{MeV} ,\quad   m_{\eta^\prime}=957.78(6)~\text{MeV} 
\end{equation}
While the state identification is unequivocal in this case, the experimental uncertainties are very small compared to the expected systematics of our modeling approach. Therefore, we choose to scale the mass uncertainty of these mesons over their mass similar to the pion ratio $\sigma_{m_\pi}/m_\pi$:
\begin{equation} \label{eta}
    \sigma_{m_\eta} = \frac{m_\eta}{m_\pi}\sigma_{m_\pi}\,, \qquad \sigma_{m_{\eta^\prime}} = \frac{m_{\eta^\prime}}{m_\pi}\sigma_{m_\pi}
\end{equation}
    
Concerning the pion decay constant, similar considerations on the isospin dependence are in order. It has been argued that isospin symmetry breaking brings only a second order correction in chiral perturbation theory, such that $f_{\pi^{\pm}}\approx f_{\pi^0}$ should be a good approximation~\cite{rosner2019leptonic}. Unfortunately, leptonic decay constants cannot be measured directly from experiments as only their product with CKM matrix elements appear in their (measurable) weak decay width. Therefore, their determination must rely on some additional theoretical framework (typically LQCD together with chiral perturbation theory). From Ref.~\cite{rosner2019leptonic} the theoretical preferred value for $f_\pi$ is:
    \begin{equation}
        f_\pi=\frac{1}{\sqrt{2}}(130.2 \pm 1.7) \approx 92.1 \pm 1.2~\text{MeV}
    \end{equation}
The factor $1/\sqrt{2}$ is due to a different choice of normalization to that of the NJL literature. To take systematic uncertainties into account, the recommended uncertainty was inflated by 10\%.

Surprisingly, the quantity $f_K/f_\pi$ is not very well reproduced by the NJL model~\cite{Klevansky}. Anticipating our results, the predicted values of the ratio lie typically between 1 and 1.1, while the experimental value of Ref.~\cite{rosner2019leptonic} is ${f_K/f_\pi = 1.1928(26)}$. Such a discrepancy indicates a large systematic error coming from the model for this quantity that is not well understood. To still favor larger values of this ratio, we increase the systematic uncertainty to $\sigma_{f_K/f_\pi}=0.075$.


The chiral condensate $\Sigma = -\langle \overline{q}q\rangle^{\frac{1}{3}}$ is another vacuum parameter that cannot be accessed easily by experiments. We must therefore rely again on ab initio approaches as LQCD or QCD sum rules (QCDSR) to obtain estimates. Early estimations from LQCD with two flavors \cite{lattice1999} led to the result:
    \begin{equation}
        \Sigma= 231 \pm 11 ~\text{MeV}
    \end{equation}
at a renormalization scale of 1 GeV. Estimations for the QCDSR, on the other hand, suggest the wider range \cite{Buballa,QCDSR1998}:
    \begin{equation}
         \Sigma= 229^{+33}_{-36} ~\text{MeV}
    \end{equation}
Today, the ranges recommended by the Flavour Lattice Data Group (FLAG) for two and three flavors are \cite{FLAG2019}:
    \begin{equation} \label{lqcdsigma}
          \Sigma = 266\pm10 ~\text{MeV} ~~(\text{2f})\,,\quad\Sigma= 272\pm5 ~\text{MeV} ~~(\text{3f}), 
    \end{equation}
which are larger than the previous estimations.
These results were however obtained in the chiral limit (vanishing $m_u$ and $m_d$) and at the renormalization scale of 2~GeV (which is conventional in LQCD simulations). 

With all these conflicting results, it is difficult to find a proper restriction on the chiral condensate $\Sigma$. It seems to be reasonable to exclude models predicting chiral condensates larger than ${\Sigma_{{\rm max}}\approx 275~}$MeV and smaller than $\Sigma_{{\rm min}}\approx$ 210~MeV. Therefore, in the conservative approach we implemented a Gaussian passband filter to constrain the chiral condensate in this range:
\begin{equation}\label{wsigma}
        w_{\Sigma}=\frac{1}{4}\Big[1-\erf\Big(\frac{\Sigma-\Sigma_{{\rm max}}}{\sqrt{2}\sigma_{\Sigma}}\Big)\Big]\Big[1+\erf\Big(\frac{\Sigma-\Sigma_{ {\rm min}}}{\sqrt{2}\sigma_{\Sigma}}\Big)\Big] \ ,
\end{equation}
with an estimate of the standard deviation $\sigma_{\Sigma} = 10$~MeV.
    
For all baryons of the octet, we have naturally very precise measurements of their experimental masses from the PDG. Just like for mesons, the isospin partners are degenerate in our approximation and we determine the mass of the isospin-averaged state (that we will call nucleon, $\Lambda$, $\Sigma$ and $\Xi$ baryons)  by calculating the barycenter of the masses of the different physical states. Clearly, the uncertainties coming from the isospin mass splittings ($\approx1~$MeV for the nucleon, $\lesssim 10~$MeV for the $\Sigma$ and $\Xi$ baryons) are very small. 
The difficulty lies in the estimation of the systematic uncertainty of our formalism for the construction of baryons, that relies on several nested approximations (NJL model, mean field approach, quark-diquark reduction of the 3-body equation and static approximation). To be on the conservative side we choose to fix the nucleon mass uncertainty with ${\sigma_{m_N}=50\times(m_n-m_p)}\approx60~$MeV, and scale this uncertainty to the other octet baryons with the same method, as we did for $\eta$ mesons in Eq.~(\ref{eta}).
    
In the case of the $\Delta$ baryon, the PDG database reports a Breit-Wigner mass $m_\Delta = 1232\pm2$~MeV and a full width $\Gamma_\Delta = 117\pm3$~MeV. The masses of the individual $\Delta^{++}$, $\Delta^+$, $\Delta^0$ and $\Delta^-$ resonances are not known very precisely. Since the decay of the $\Delta$ baryon is not predicted within our model, we suggest that the width of the resonance is a good indicator of the scale of precision that cannot be accounted for by the NJL model, such that we choose $\sigma_{m_\Delta}=\Gamma_\Delta$. For the remaining baryons of the decuplet ($\Sigma^*$, $\Xi^*$ and $\Omega$) we determine again the isospin symmetric mass by averaging the physical masses from the PDG database, and scale $\sigma_{m_\Delta}$ to the mass of the corresponding baryon in a similar fashion to Eq.~(\ref{eta}).


\begin{table*}[htbp]
\centering
\begin{tabular}{|c|c|c|c|c|c|c|c|c|}
\hline
 & $m_\pi$ & $f_\pi$ & $m_K$ & $f_K/f_\pi$ & $m_\eta$ & $m_{\eta^\prime}$& $-\langle\overline{q}q\rangle^{1/3}$&\\
\hline
$\overline{o}$ &  138.04 & 92.07 & 495.64 & 1.1928 & 547.862 & 957.78& 210 - 275 &\\
$\sigma$ & 4.6 & 1.3 & 3.93 & 0.075 & 18.23 & 31.87& 10& \\
\hline
\hline
& $m_N$ & $m_\Lambda$ & $m_\Sigma$ & $m_\Xi$ & $m_\Delta$ & $m_{\Sigma^\star}$&$m_{\Xi^\star}$&$m_{\Omega}$\\
\hline
$\overline{o}$ &  938.92 & 1115.683 & 1193.15 & 1318.28 & 1232 & 1384.57&1533.40&1672.45\\
$\sigma$ & 64.66 & 76.84 & 82.18 & 90.79 & 117 & 131.49 & 145.62 & 158.83\\
\hline
\end{tabular}
\caption{Mean and standard deviation used for the Gaussian weights associated with each quantity. All quantities (except $f_K/f_\pi$ which is dimensionless) are given in units of MeV.}
\label{PDGspectrum}
\end{table*}

\section{Results and discussion} \label{sec::discussion}

\begin{figure*}[htbp]
    \centering
    \includegraphics[scale=0.17]{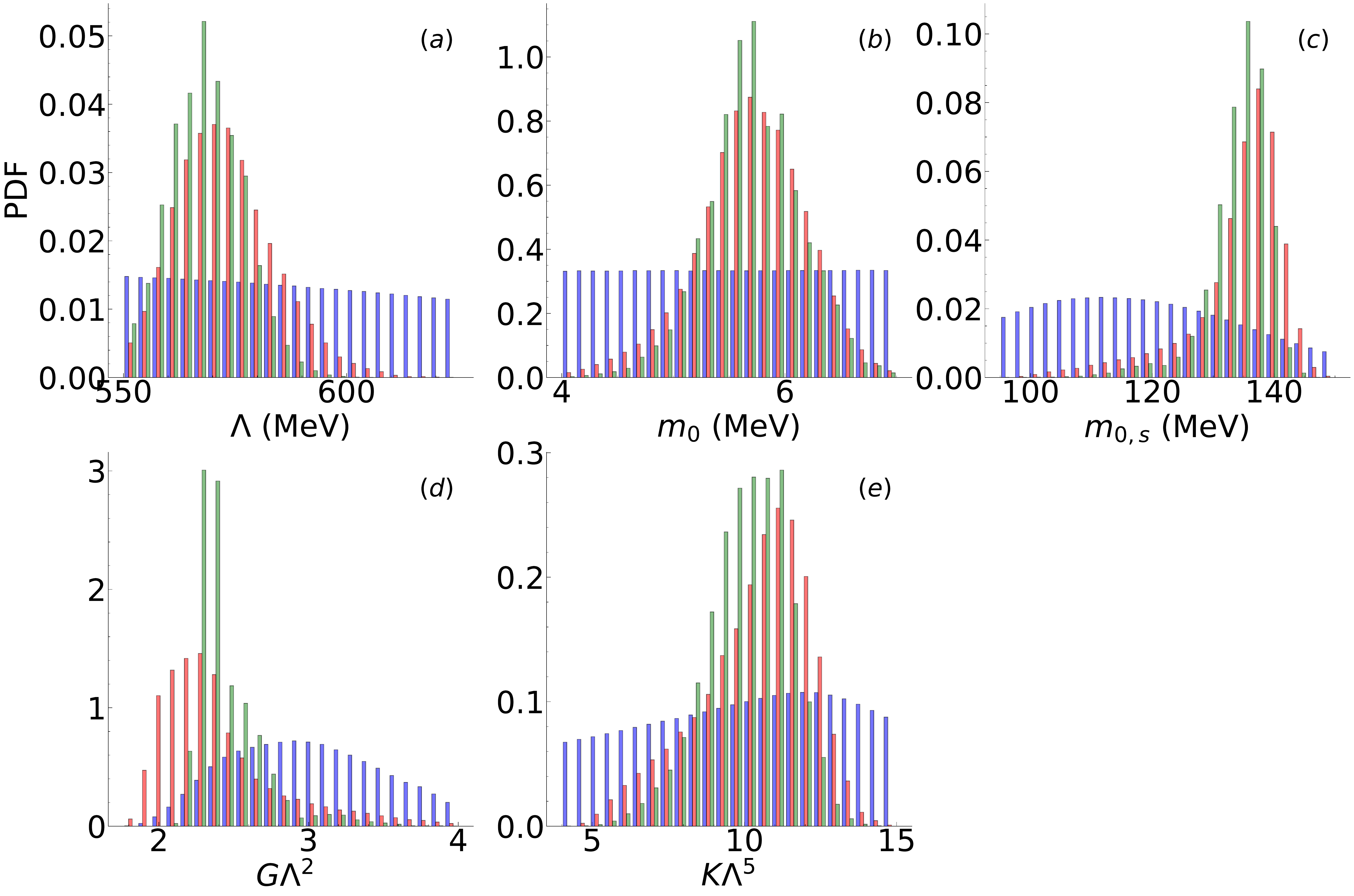}
    \caption{Posterior distributions of the SU$(3)_f$ NJL model parameters with weights W1 (blue), W2 (red) and W3 (green); see text for details. All histograms are normalized to unit area.}
    \label{params3f}
\end{figure*}

To compare the effects brought by each weight in the likelihood, they were gathered in three weighting procedures that are then compared on the following figures:

\begin{itemize}
    \item {\bf W1 (represented in blue)}: No weight is assigned to the data, i.e. $\mathcal{P}( {\rm data}|\{o_i\})=1$ according to Eq.~(\ref{eq::likelihood}). The distribution simply reflects the natural predictions of the model in the explored prior parameter space. However, one should keep in mind that the additional conditions on the stability of the bound states (see Eq.\ref{stable} and \ref{baryonstab}), which may seem minor at first, already affect the model selection process.
    \item {\bf W2 (represented in red)}: All weights associated with the properties of the meson octet (including $f_\pi$ and $f_K/f_\pi$), as well as the chiral condensate filter are implemented.
    \item {\bf W3 (represented in green)}: The weights associated with each baryon of the octet and decuplet are added on top of the ones already in W2. 
\end{itemize}


In this section, we will compare the posterior probability distribution functions (PDFs) conditioned by each weighting procedure W1, W2 and W3, in order to reveal the role that is played by the different constraints. As a summary of our results, we provide in App.~\ref{Appsummary} a table collecting the predicted mean values and uncertainties associated with each quantity examined in this paper for the weighting W3, that is when all possible physical constraints have been applied.

\subsection{Model parameters}

\begin{figure}[htbp]
    \centering
    \includegraphics[scale=0.124]{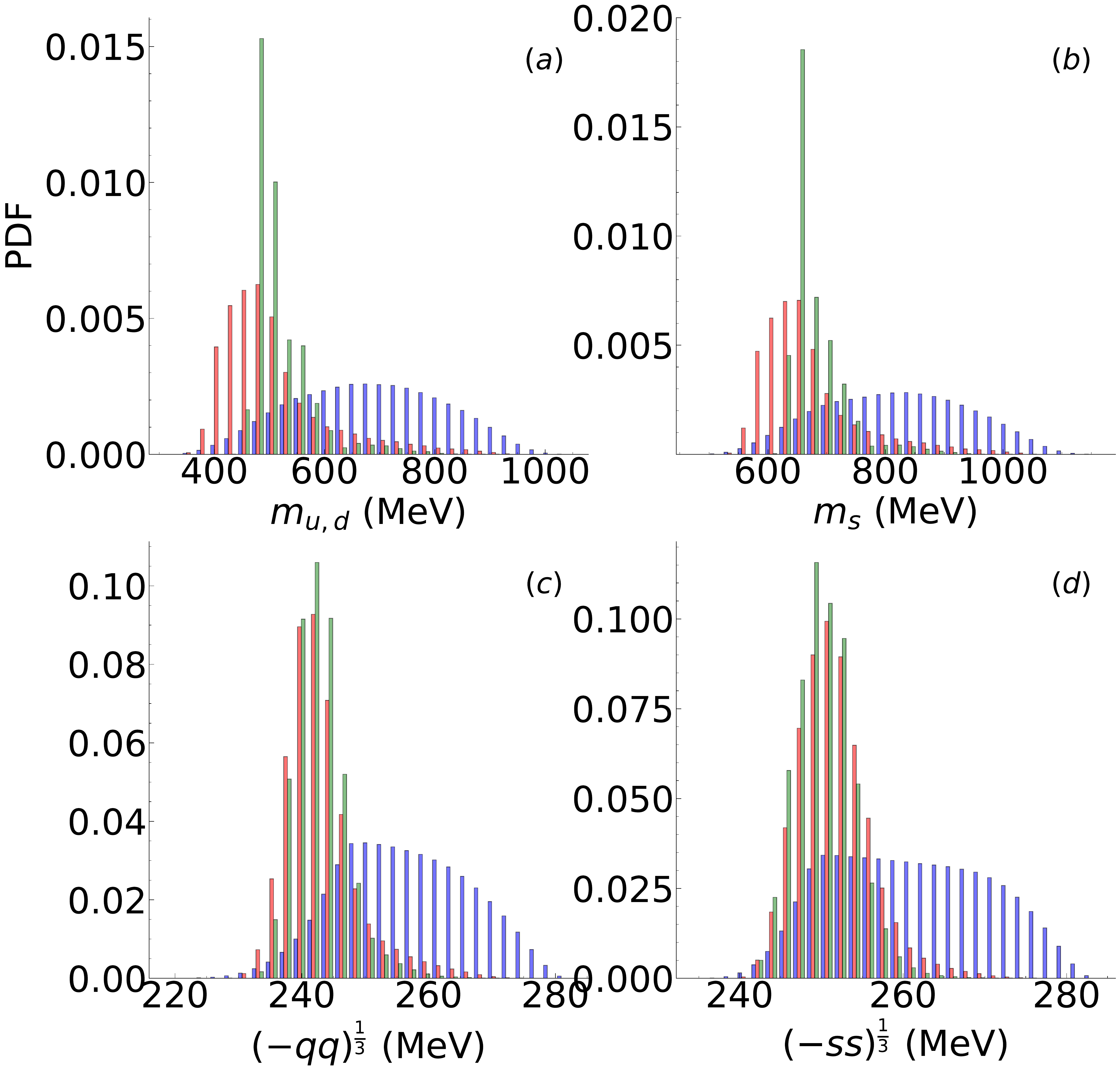}
    \caption{Posterior distributions of the quark mass and chiral condensates for light and strange flavors with weights W1 (blue), W2 (red) and W3 (green) (see main text for details). All histograms are normalized to unit area.}
    \label{quarks3f}
\end{figure}

In Fig.\ref{params3f}, we show the PDFs of the five model parameters (leaving the discussion of the diquark parameters to Sec.\ref{sec::baryonres}). As expected, if no weights are assigned to the data (W1), the probability distributions remain almost flat, reflecting our use of a flat prior for these parameters. The deviations from a flat distribution (particularly notable for the quantity $G\Lambda^2$) epitomize the effect of selecting models that are able at least to produce a stable spectrum of mesons and baryons. Low values of $G\Lambda^2$ ($\lesssim 2$) typically lead to relatively small quark masses, which may damage the stability of baryons due to Eq.~(\ref{baryonstab}). Once we enforce physical values for the hadron spectrum (W2 and W3), the distributions of all parameters become peaked around a preferred value, which validates our choice for the prior. The parameter set employed in a previous work using the same set of approximations is in good agreement with these results \cite{TorresRincon2015}.
We should mention that we also checked with larger priors that there is no statistically significant increase of these PDFs outside of the parameter ranges used in this work.

The PDFs of the dressed quark masses and chiral condensates are presented in Fig.\ref{quarks3f}. The light quark mass is found to lie between 315~MeV and 1015~MeV within our parameter prior. Once the W2 filter on meson masses is implemented, this range is narrowed down to $m_{u,d}\lesssim600~$MeV, with a preference around 430~MeV. The reproduction of baryon masses disfavors smaller values and narrows down further the light quark mass to $m_{u,d}=515~\pm53$~MeV, although W2 and W3 remain largely compatible.
A similar pattern is observed in the strange sector. Such an effect can be explained by the requirement of a finite binding energy for both diquarks and baryons as we shall discuss below.
In addition, it is interesting to notice that values close to the naive estimate ${m_{u,d} \approx m_N/3 \approx 313~}$MeV are rejected already by W1. This implies that low quark masses are not compatible with the reproduction of the full hadron spectrum, as long as the stability conditions~(\ref{stable}) and ~(\ref{baryonstab}) are satisfied. This is in agreement with the findings of Ref.~\cite{Buck:1992wz}.

The light chiral condensate $\Sigma=-\langle \overline{q}q\rangle^{\frac{1}{3}}$ is found to lie in the range {218$~$MeV$\,<\Sigma<\,$282~MeV} in our prior. This is already in good agreement with previous empirical knowledge on this quantity, implying that the effects of the weight $w_\Sigma$~ (\ref{wsigma}) will be overall very small.
Imposing the matching of the meson spectrum only gives a small range of compatible values ${\Sigma = 243 \pm 6~}$MeV, which is not affected much by the additional conditions of W3 on baryons, contrary to the quark mass distribution. This result is quite small compared to the estimations from LQCD (see Eq.~\ref{lqcdsigma}).

\begin{figure}[htbp]
    \centering
    \includegraphics[scale=0.124]{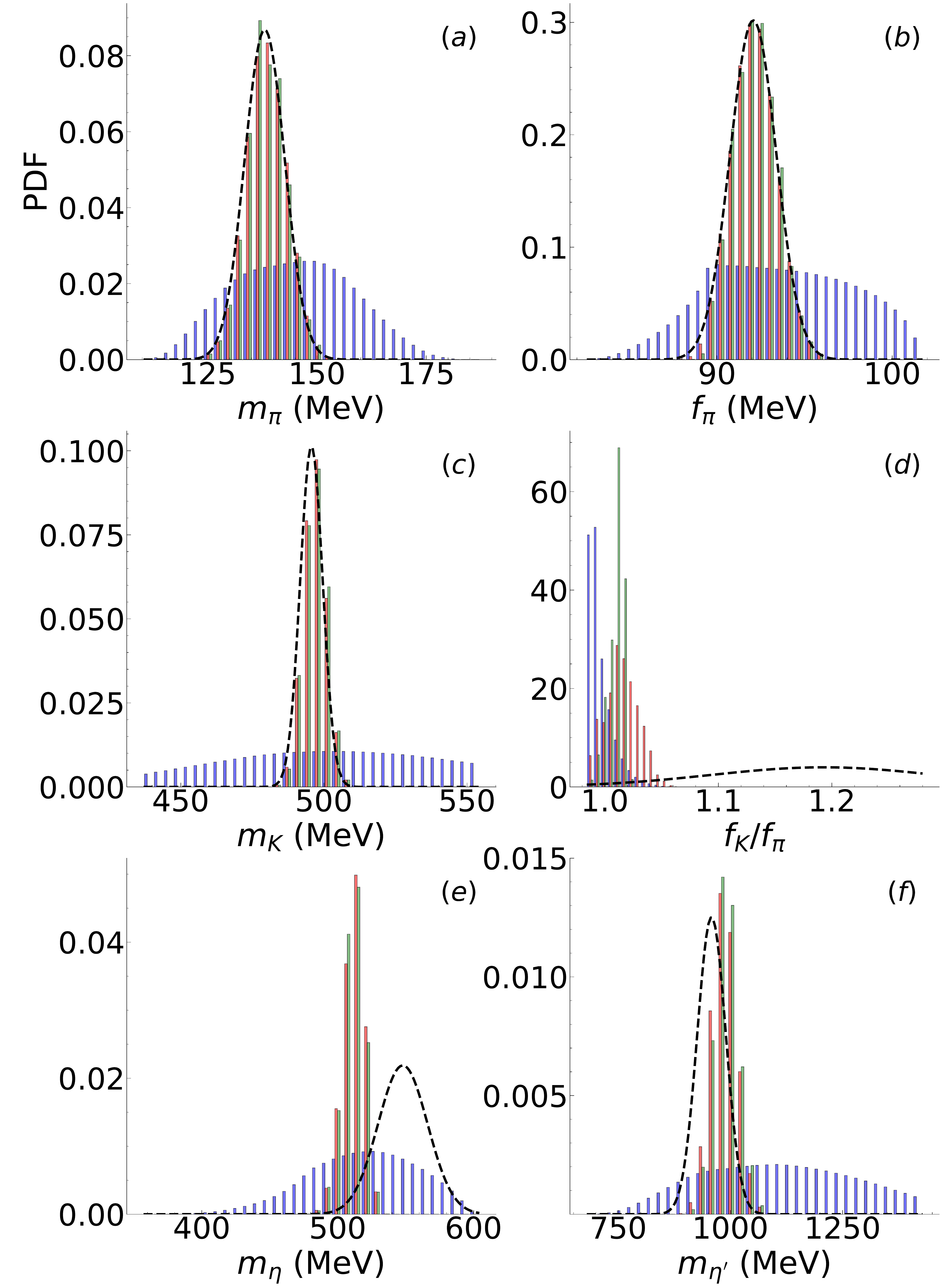}
    \caption{Posterior distributions of the meson octet masses and decay constants with weights W1 (blue), W2 (red) and W3 (green) (see main text for details). The dashed lines indicate the likelihood weights implemented in W2 and W3 for each quantity. All histograms are normalized to unit area.}
    \label{mesons3f}
\end{figure}

We show in Fig.~\ref{mesons3f} the PDFs of the observables for the pseudo-scalar octet, \textit{i.e.} meson masses and decay constants. For each quantity, we also show with a dotted line the weight distribution of the likelihood illustrating the constraints implemented in both W2 and W3. In the case of the pion mass, pion decay constant and kaon mass, the posterior distributions of W2 and W3 follow very well the likelihood, such that the uncertainties on these quantities are accurately reproduced. On the other hand, the likelihood on the ratio $f_K/f_\pi$ is poorly reproduced, as the model is known to be unable to produce values larger than $\approx 1.05$ for this ratio~\cite{Klevansky}, unless a very small quark mass is adopted (see e.g.~\cite{Ninomiya:2014kja}). 
For the $\eta$ and $\eta^\prime$ mesons, we see that while the physical values for the masses of both states are able to be reached by the model, they unfortunately cannot be simultaneously accounted for. Therefore, when trying to enforce physical values, one needs to make a compromise with slightly larger $m_\eta^\prime$ and smaller $m_\eta$. 
Both of these failures could be indicators of the limits of the NJL model as well as the mean field and ring approximation used to treat the problem. It also shows the limits of a strategy of  direct fitting of the model parameters to the data. For instance, the 't\,Hooft coupling $K$ is often used in order to fine tune the $\eta-\eta'$ mass splitting, but it turns out this cannot be done without damaging the quality of the fit to other quantities. Treating all experimental quantities on equal footing (factoring the uncertainties), we guarantee to find the best middle ground the NJL model can offer.

\subsection{Predictions for the baryon spectrum}\label{sec::baryonres}

\begin{figure}[htbp]
    \centering
    \includegraphics[scale=0.124]{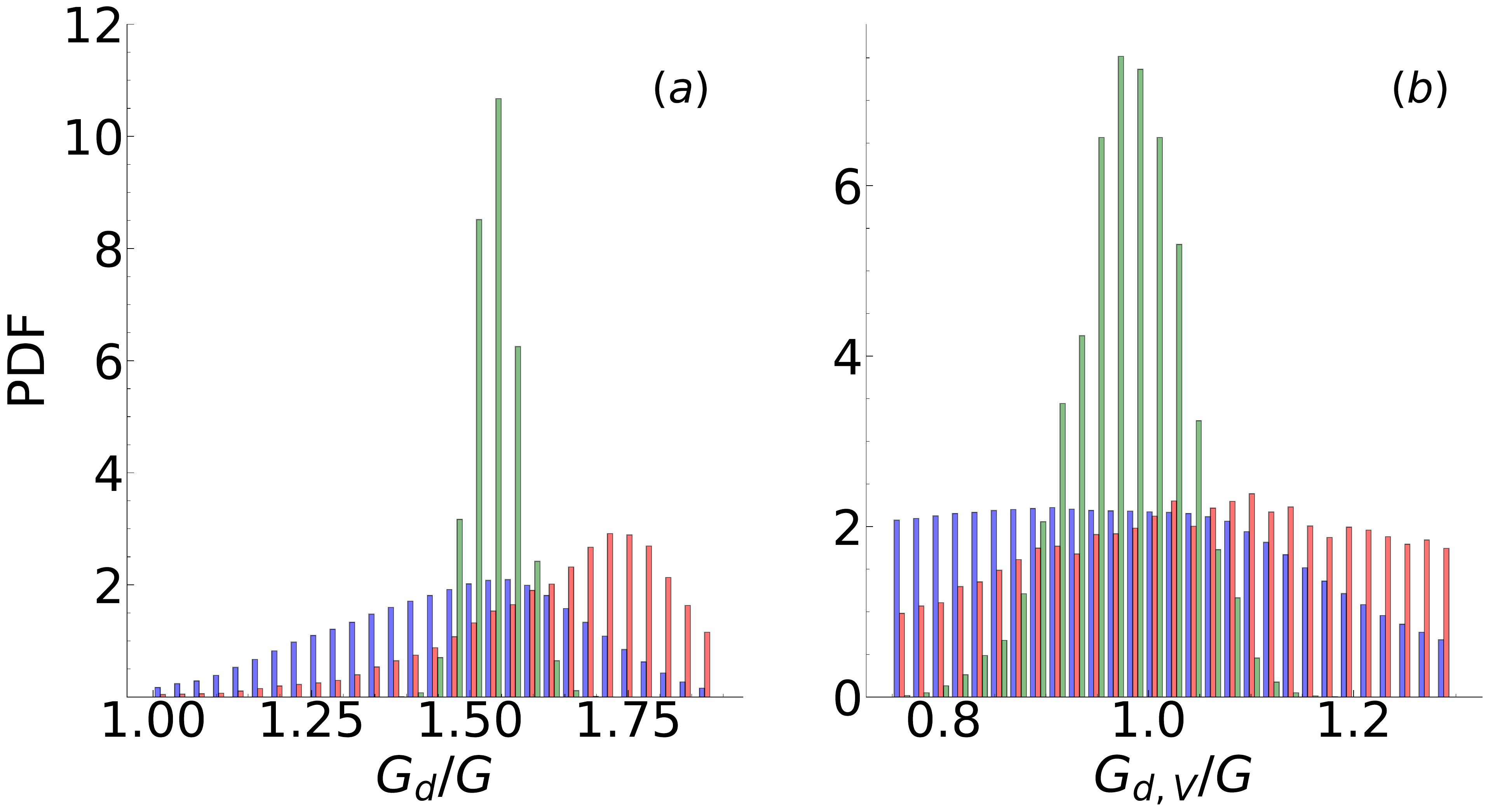}
    \caption{Posterior distributions of the scalar diquark coupling (a) and axial-vector diquark coupling (b) with weights W1 (blue), W2 (red) and W3 (green) (see main text for details). All histograms are normalized to unit area.}
    \label{paramsdi3f}
\end{figure}

In Fig.~\ref{paramsdi3f}, we show the prior and posterior PDFs of the diquark coupling ratios $G_d/G$ and $G_{d,V}/G$. In both cases, we see that relatively large coupling ratios are needed to sustain the existence of the baryon states. The meson and baryon constraints altogether (W3) select rather precise values for the ratios, $G_d/G = 1.53\pm0.04$ and $G_{d,V}/G=0.97\pm0.06$. These results are surprisingly far from those obtained via the Fierz formulas, which suggest $G_d/G = 3/4$ and $G_{d,V}=3/8$, respectively. This confirms that one should take caution when applying these relationships within a NJL-like model. This value of the diquark coupling $G_d/G$ is also typically larger than the one used in the literature to study the transition to color superconducting matter at high density \cite{Sedrakian,Kl_hn_2013,Blaschke2021,Contrera:2022tqh}.

Even though the quantity $G_d/G$ affects only the diquark and baryon sectors, we can also observe in Fig.\ref{paramsdi3f}a a strong preference towards large values with weighting W2 (which remains agnostic on the baryon masses). This is explained by the preference of W2 towards smaller quark masses, as baryons require stronger binding to remain stable in this case.

\begin{figure}[htbp]
    \centering
    \includegraphics[scale=0.124]{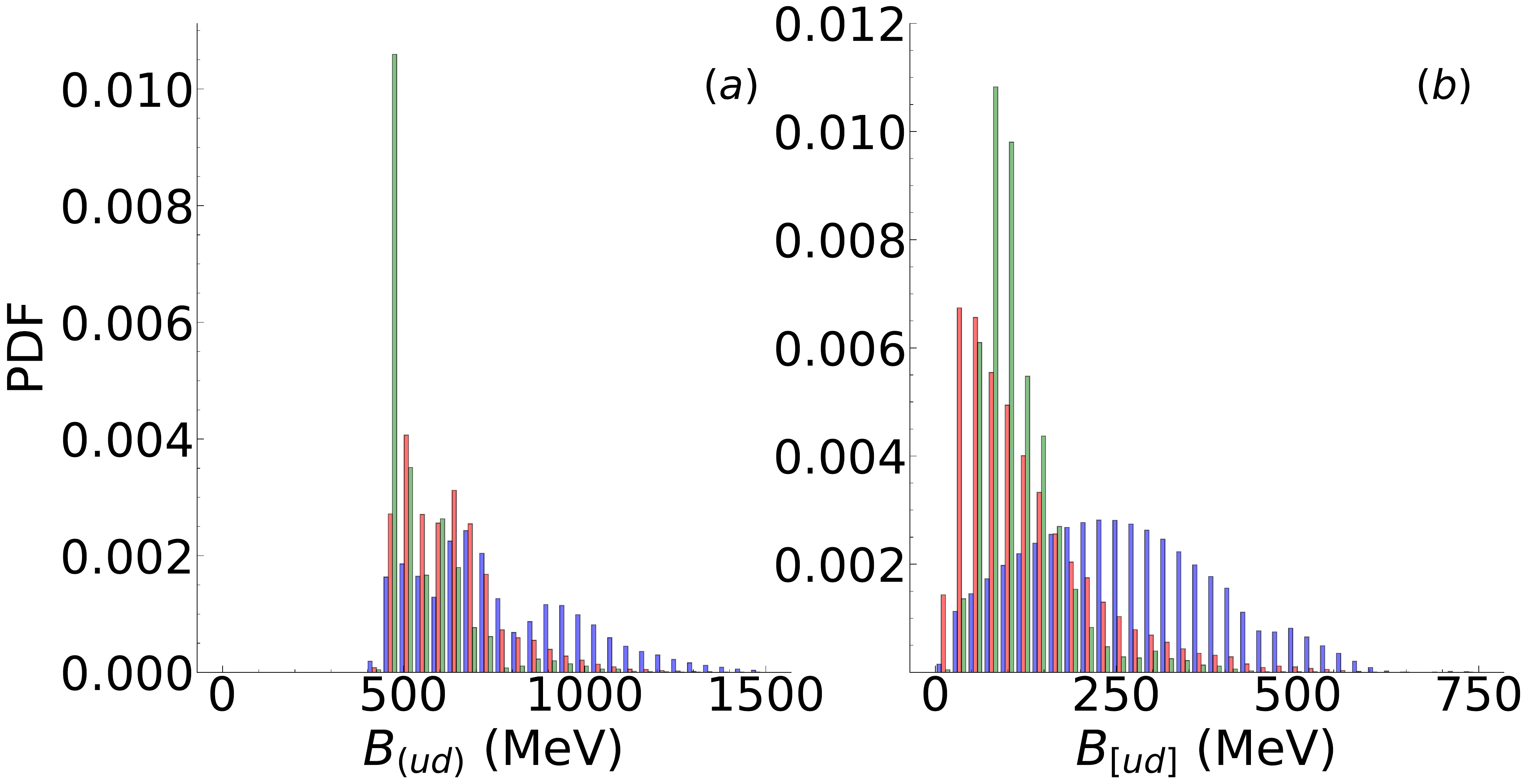}
    \caption{Posterior distributions of the scalar diquark binding energy (left) and axial-vector diquark binding energy (right) with weights W1 (blue), W2 (red) and W3 (green) (see main text for details). All histograms are normalized to unit area.}
    \label{binding_ud3f}
\end{figure}

In Fig.~\ref{binding_ud3f} we show the posterior distributions of the binding energies $B_d = 2m_{q} - m_d$ for both of the non strange scalar and axial diquarks.
We can observe a striking difference of behavior between the two species; while the axial diquark is allowed to be very weakly bound within the NJL model, this is not the case for the scalar diquark, whose binding energy is bounded to be larger than $B_{(ud)}^{min}\approx$ 500 MeV. This behavior can be explained by the fact that even with W1, our description of the nucleon requires a very large diquark coupling to be bound, which prevents the diquark to be weakly bound. This effect is not present for the axial diquark because of its larger mass and the lower value of the diquark-vector coupling required to fit the physical $\Delta$ mass. We observed a similar behavior for the other scalar and axial diquarks in the strange sector. It should be stressed that this feature is not a consequence of our optimization of the parameters to reproduce experimental results, but is inherent to the NJL model (at least within our approximations).

This property explains why our assumption of the existence of stable baryons (and the nucleon in particular) requires a sufficiently large quark mass. Indeed, if we rewrite the stability conditions (Eq.(\ref{stable}) and (\ref{baryonstab})) in the form:

\begin{equation}    \label{nucbinding}
3m_{u/d} = m_N + B_{(ud)} + B_N,
\end{equation}
where ${B_N = m_{(ud)}+m_{u/d}-m_N\geq 0}$ is the nucleon binding energy, it becomes apparent that the minimum possible value for the quark mass is ${m_{u/d} = (m_N+B_{(ud)}^{min})/3\approx 480~}$MeV when $m_N$ is fixed to its physical value. 
\begin{figure}[htbp]
    \centering
    \includegraphics[scale=0.124]{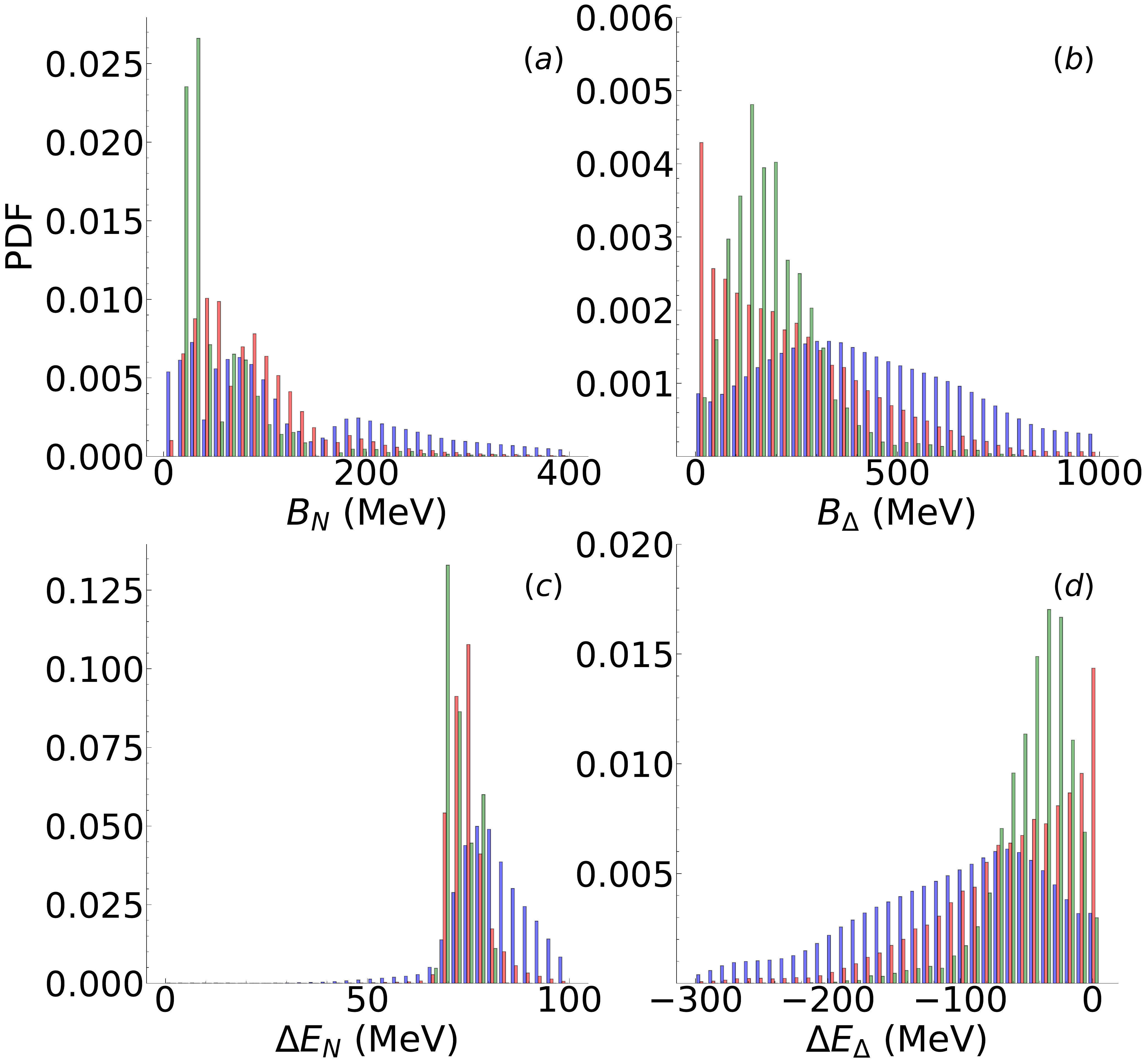}
    \caption{Posterior distributions of the baryon binding energy ${B_B = m_d+m_q-m_B}$ and energy per quark difference ${\Delta E_B = m_B/3 - m_d/2}$ for the nucleon (left column) and the $\Delta$ baryon (right column). All histograms are normalized to unit area.}
    \label{bindingND}
\end{figure}

Finally, we compare in the top panel of Fig.~\ref{bindingND} the posterior distributions of the binding energy of the nucleon and $\Delta$ baryon. We notably see that the constraints from W3 restrict the nucleon binding energy to a maximum of about 100~MeV, while values as large as 500~MeV can be obtained for the $\Delta$ baryon. The most probable scenario is that of a very weakly bound nucleon with $B_N<30~$MeV. Since our framework describes both the nucleon and its constituent diquark with a single coupling constant $G_d$, we can expect that the diquark and nucleon binding are very heavily correlated. This lack of freedom can explain why the only possibility compatible with the physical value of $m_N$ is that of a very unbalanced repartition of the binding energies with a strongly bound diquark and a weakly bound nucleon. Because of the larger mass of the $\Delta$, both diquark and baryon can bear reasonable binding energies. This discrepancy is also related to the larger value of the scalar coupling compared to the vector one.

On the bottom panel, we also compare the distributions of the quantity $\Delta E_B = m_B/3 - m_d/2$ which represents the energy per quark difference between the baryon and the diquark. While this difference is always positive for the nucleon in our model, it is almost always negative for the $\Delta$ baryon. The same pattern is observed for every other members of the baryon octet and decuplet. This illustrates again the difference of behavior between the two baryon families in our framework; in the case of the octet, the diquark is very strongly bound and therefore very light, whereas the binding energy is well distributed between the axial diquarks and baryons of the decuplet. 
The sign of $\Delta E_B$ may have strong implications in medium due to the lack of color confinement of the model \cite{Birse1999,Pepin_2000}. The fact that it is positive for the baryon octet means that scalar diquarks are energetically favored over baryons. Since there are no confinement mechanism to prevent the diquark colored state to remain, they are allowed to survive in medium. At finite chemical potential and low temperatures, this may lead to the formation of a diquark Bose-Einstein condensate \cite{Hands:2006ve,Abuki:2010jq}.

\subsection{Correlations}

\begin{figure*}[p]
    \centering
    \includegraphics[scale=0.5]{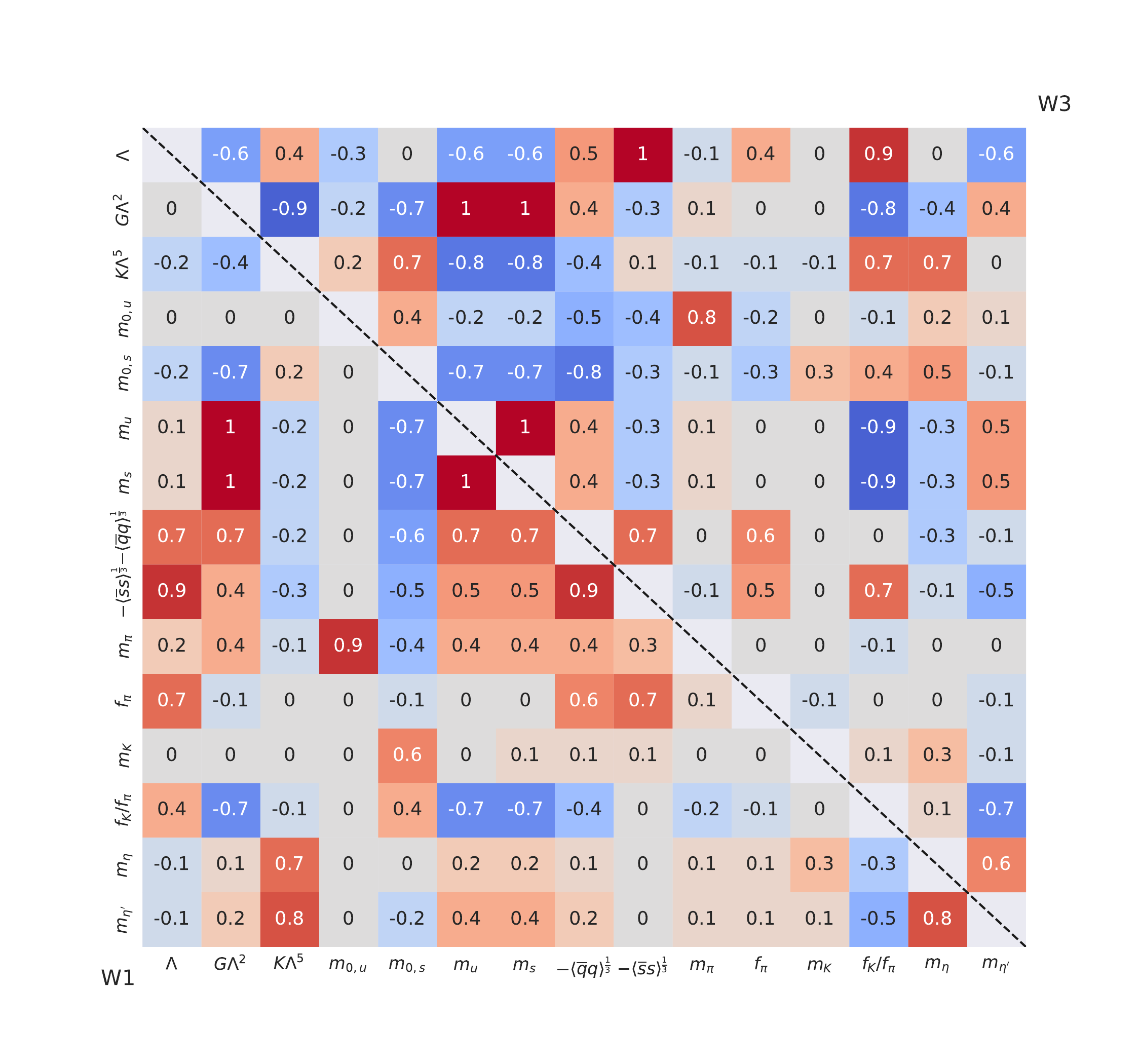}
    \caption{Correlation coefficients between the parameters and observables in the meson sector. The bottom-left half contains correlations when no weighting procedure is applied to the data (weight W1; see Sec.\ref{sec::discussion}), while the top right corner highlights the remaining correlations after all the constraints have been applied (weight W3).}
    \label{corrcombinedrun5}
\end{figure*}

\begin{figure*}[p]
    \centering
    \includegraphics[scale=0.5]{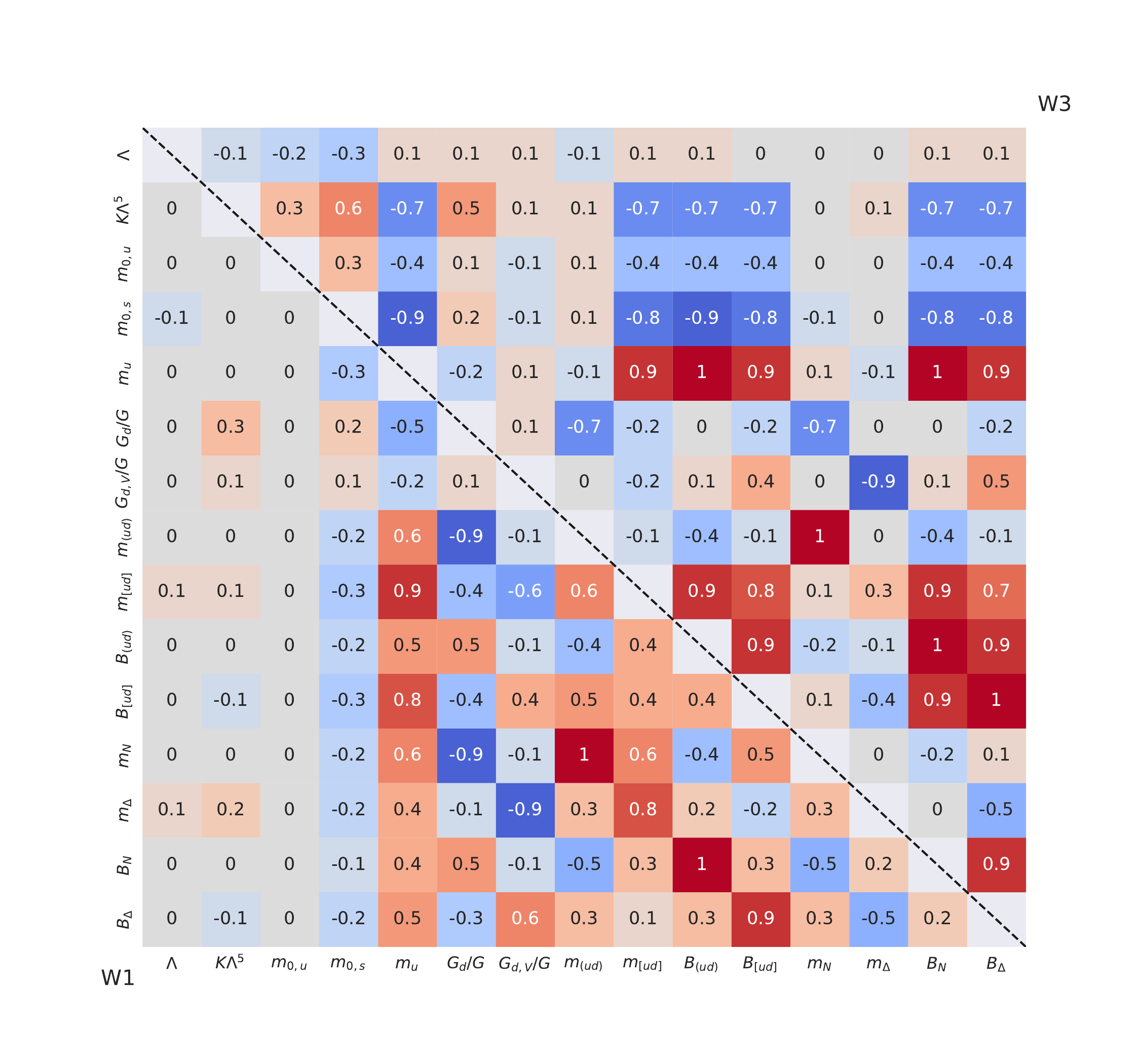}
    \caption{Correlation coefficients between the parameters and observables in the baryon sector. The bottom-left half contains correlations when no weighting procedure is applied to the data (weight W1), while the top right corner highlights the remaining correlations after all the constraints have been applied (weight W3).}
    \label{corrcombinedrun5b}
\end{figure*}

To complete our analysis of the posterior distributions, we provide correlation tables to understand the interplay on one side between the model parameters and the pseudoscalar sector in Fig.~\ref{corrcombinedrun5}, and on the other side between the model parameters and the baryon sector in Fig.~\ref{corrcombinedrun5b}. The correlation between two quantities $o_i$ and $o_j$ is interpreted through their linear Pearson coefficient $C_{ij}$ defined by

\begin{equation}    
C_{ij}=\frac{\E[o_io_j]-\E[o_i]\E[o_j]}{\sqrt{\E[o_i^2]-\displaystyle\E[o_i]^2}\sqrt{\E[o_j^2]-\E[o_j]^2}} \, ,
\end{equation}
where $\E$ is the expectation operator.
On each figure, the coefficients are shown for the probability distributions obtained with weightings W1 and W3 in the lower left and upper right triangle, respectively. This allows us to distinguish between the correlations coming from the modeling itself, and the ones created by the constraints we impose on each parameter sets. In the following, we highlight the most important conclusions that we draw from these two figures.

First, we notice that both the light and strange quark masses are very strongly correlated to the dimensionless scalar coupling $G\Lambda^2$. This is not surprising, as the scalar quartic term is responsible for the mechanism of spontaneous symmetry breaking in the theory. However, this shows that the second term of Eq.~(\ref{gap}) remains dominant against the bare mass and the 't\,Hooft term for all possible parametrizations. In addition, we conclude that the quantity $G\Lambda^2$ is the most relevant to control the value of both quark masses, which are in turn heavily correlated with each other. On the other hand, the chiral condensates (especially the strange one) are better characterized by the value of the momentum cutoff $\Lambda$. In the absence of constraints, the light bare mass only affects the pion mass, which can be easily understood from its origin as a pseudo-Goldstone boson. Once the pion mass is fixed to its physical value, the value of $m_{0,u}$ is adapted to satisfy the condition to bring $m_\pi$ to its physical value, which brings minor correlations with the masses and condensates. In a similar fashion, the quantity $K\Lambda^5$ influences the most the masses of the $\eta$ and $\eta^\prime$ mesons due to its role in breaking the U(1)$_A$ symmetry, but inherits significant anticorrelations with the quark masses once the meson masses are filtered.

Regarding the baryon sector (Fig.~\ref{corrcombinedrun5b}), we observe, as expected, strong correlations between the diquark and baryon sectors. In particular, in the case of W1, the masses of the baryons are strongly determined by the mass of the corresponding diquark, and to a lesser extent by the quark mass. In addition, strong a anticorrelation is observed between the diquark couplings which are naturally proportional to the baryon binding energy. After matching with the physical meson and baryon masses, the role of the quark mass $m_u$ becomes much more significant, the binding energies {$B_{(ud)}$, $B_{[ud]}$, $B_{N}$} and $B_\Delta$ having Pearson coefficients close to unity with $m_u$. These correlations are easy to understand from Eq.~(\ref{nucbinding}) (and its equivalent for the $\Delta$~baryon), where the relationship between the total binding energy and the quark mass appears to be linear assuming the bound state mass to be fixed. This emphasizes the central role of the quark mass as an energy scale in the problem, as its value remains not very well constrained by the physical input (see Fig.\ref{quarks3f}).
This correlation with the quark mass is also very strong for the mass of the axial diquark~$m_{[ud]}$, but surprisingly disappears completely for the scalar diquark mass~$m_{(ud)}$ once the nucleon mass is fixed. We expect that this effect is related to the different behavior of the scalar diquark observed in Sec.~\ref{sec::baryonres} as well as the larger value of the diquark coupling required in the scalar sector to match the mass of the baryon octet, which prevents large deviations of~$m_{(ud)}$ from its average value. This difference is also illustrated by the larger sensibility around the physical point of the nucleon mass towards the mass of its diquark component ($C_{ij}\approx1$) than for the $\Delta$ baryon ($C_{ij}\approx0.3$).

Note that just like in Sec.~\ref{sec::baryonres}, we only discuss the correlation coefficients related to the nucleon and $\Delta$ baryon (together with their corresponding diquark), as we observed a very similar correlation pattern for the other baryons belonging to the same multiplet.  

\section{Summary} \label{sec::summary}

In the present work, we have performed a Bayesian analysis of the properties of the meson and baryon spectra as predicted by the SU$(3)_f$ Nambu--Jona-Lasinio model. The Bethe-Salpeter equations in the RPA approximation as well as the Faddeev equations in a quark-diquark static approximation were solved to compute the masses of the lowest mesonic and baryonic states for a large variation within the parameter space. The results were then confronted with the experimentally measured masses of the physical states to select plausible models.

As it has been shown by previous studies, we confirm that the NJL model is able to satisfactorily describe the main traits of QCD phenomenology in the vacuum, including the spectrum of the pseudoscalar meson octet and of the baryon octet and decuplet. Nevertheless, some discrepancies were observed for the $\eta$ and $\eta'$ masses as well as the kaon/pion decay constant ratio $f_K/f_\pi$. This global view is new in our approach, as in previous work the parameter set had been chosen to describe a limited and arbitrary set of experimental observables.

These results highlight the importance of taking into account the maximum number of constraints to provide a good parametrization of the model. Neglecting some of the physical input or the uncertainties associated with the measurement and modeling of the quantities described may lead to misconceptions regarding the predictive power of the model and overlook the subsequent uncertainties on the model parameters and predictions. In addition, our Bayesian framework allows us to inspect the correlations between parameters and predictions to better understand the interplay between each quantity.

In order for the predicted baryon masses to be in accordance with the physical masses, we find that large values of the scalar and vector diquark couplings must be required (${G_{d}/G = 1.53\pm0.04}$ and ${G_{d,V}/G = 0.97\pm0.06}$). In the scalar case, we find that the diquarks are systematically strongly bound (${B_{(ud)}\gtrsim 500}$~MeV), which in turn also imposes the quark mass to be sufficiently large (${m_{u/d}\gtrsim 480~}$MeV). In consequence, the average binding energy of a quark in the nucleon is always smaller that its average binding energy in the diquark which is a constituent. These findings could possibly have significant impact on the phenomenology of dense quark matter in compact stars, where the strength of the diquark coupling could lead to a strong effect of quark pairing and color superconductivity. This relies however on the assumption that the large diquark couplings could be maintained in-medium, as the density dependences of the couplings are very much uncertain.

\acknowledgements
This study is part of a project that has received funding from the European Union’s Horizon 2020 research and innovation program under grant agreement STRONG – 2020 - No 824093.
The work of J.M.T.-R. is supported by Project no. PID2020-118758GB-I00, financed by the Spanish
MCIN/ AEI/10.13039/501100011033/, and Project no. 315477589 - TRR 211 (Strong-interaction matter under extreme conditions), financed by the Deutsche Forschungsgemeinschaft.

\addcontentsline{toc}{section}{Appendix}
\section*{Appendix}
\setcounter{section}{0}
\setcounter{subsection}{0}
\setcounter{figure}{0}
\renewcommand{\thesubsection}{\Alph{subsection}}
\renewcommand{\thefigure}{\Alph{subsection}.\arabic{figure}}
\setcounter{equation}{0}
\numberwithin{equation}{subsection}

\subsection{One-quark loop integral~\label{app:I1}}

The loop integral $I_1$ appears in the calculation of the chiral condensates Eq.~(\ref{chircond2}) and it is given by,
\begin{equation}
    I_1(m,\Lambda)=  \int^\Lambda\frac{d^4k}{(2\pi)^4}\frac{1}{k^2-m^2+i\varepsilon} \ . 
\end{equation}
In the 3D noncovariant momentum cutoff scheme, one obtains after applying Cauchy's integral formula,

\begin{equation}
\begin{split}
    i I_1(m,\Lambda) &= \int^\Lambda\frac{d^3k}{(2\pi)^3}\frac{1}{2E_k}\\
     & = \frac{\Lambda^2}{8\pi^2}\Big[\sqrt{1+\frac{m^2}{\Lambda^2}}-\frac{m^2}{\Lambda^2}\log\Big(\frac{\Lambda+\sqrt{\Lambda^2+m^2}}{m}\Big)\Big] \ ,
     \end{split}
\end{equation}
where $E_k=\sqrt{ {\bf k}^2+m^2}$. This expression can be used to solve the self-consistent gap equation~(\ref{gap}) in order to find the vacuum quark masses.

\subsection{Meson and diquark polarization functions} \label{Appendixpol}

We start by recalling the general expression of the loop function associated with the bound state $X$,
\begin{equation}
    \Pi_X(p)=i\int^\Lambda\frac{d^4k}{(2\pi)^4}\Tr\Big(\Gamma_X S(p+k) \Gamma_X^\dagger S(k)\Big) \ ,
\end{equation}
where the dressed fermion propagator $S(k)$ is given by the Eq.~(\ref{quarkpropagator}).
All the interaction vertices $\Gamma_X$ in spinor, flavor and color space associated with each bound state are gathered in Table~\ref{vertices}. Pseudo-scalar mesons have negative parity, yielding a factor $i\gamma_5$ in the spinor space. They are all colorless states, so we select the singlet color representation. 
For the pion and kaon, since we work under the isospin approximation one can choose in flavor space any representant of the pion triplet ($\pi^+$, $\pi^-$, $\pi^0$) and kaon quadruplet ($K^+$, $K^-$, $K^0$, $\overline{K^0}$); here we choose the $\pi^0$ and $K^0$ channels for illustrative purposes. For the $\eta$ and $\eta^\prime$ mesons, a diagonalization in the flavor subspace (0,8) is necessary since the polarization matrix has non-zero off-diagonal terms. We refer to Ref.~\cite{Rehberg_su3} for the technical details on this procedure (see also Ref.~\cite{Blanquier2011} for an extension without the isospin approximation).

In the diquark case, scalar diquarks are needed for the construction of the baryon octet, while axial diquarks are required to describe the baryon decuplet. Their spinor channel includes an additional $C$ matrix in order to select a particle-particle scattering channel. 
Scalar diquarks belong to the antisymmetric flavor antitriplet $\overline{\bm{3}}_f \subset \bm{3}_f\otimes\bm{3}_f$. 
In the isospin approximation, we can choose $(us)$ as a representative of the degenerate pair together with $(ds)$. 
Axial diquarks belong to the symmetric flavor sextet ${\bm 6}_f \subset {\bm{3}_f\otimes\bm{3}_f}$. We choose $[ud]$ as the representative of the isospin triplet {($[ud]$, $[uu]$, $[dd]$)} and $[us]$ as the representative of the degenerate pair together with $[ds]$.
In order to produce colorless baryons, diquarks must belong to the antisymmetric color antitriplet representation of ${\bm{3}_c\otimes\bm{3}_c}$.
They are therefore colored objects with antisymmetric color wave functions $|rb-br\rangle$, $|gb-bg\rangle$ or $|rg-gr\rangle$. 
The color triplets are of course degenerate, and are combined together with quarks of the different colors to form colorless baryons.

\begin{table}[htp]
\centering

\begin{tabular}{|c|c|c|c|}

\hline
$X$ & Spinor & Flavor & Color \\
\hline
$\pi$&  $i\gamma^5$ & $\tau_3$& $\mathbbb{1}_c$  \\
$K$&  $i\gamma^5$ & $\frac{1}{\sqrt{2}}(\tau_6-i\tau_7)$ & $\mathbbb{1}_c$   \\
$\eta$&  $i\gamma^5$ & $(\tau_0,\tau_8)$& $\mathbbb{1}_c$   \\
$\eta^\prime$&  $i\gamma^5$ & $(\tau_0,\tau_8)$& $\mathbbb{1}_c$ \\
$(ud)$&  $Ci\gamma^5$ & $-\tau_2$& $-\lambda_2$/$\lambda_5$/$-\lambda_7$   \\
$(us)$&  $Ci\gamma^5$ & $\tau_5$& $-\lambda_2$/$\lambda_5$/$-\lambda_7$ \\
$[ud]$&  $C\gamma^\mu$ & $\tau_1$& $-\lambda_2$/$\lambda_5$/$-\lambda_7$  \\
$[us]$&  $C\gamma^\mu$ & $\tau_4$&$-\lambda_2$/$\lambda_5$/$-\lambda_7$  \\
$[ss]$&  $C\gamma^\mu$ & $\frac{1}{\sqrt{3}}\tau_0-\sqrt{\frac{2}{3}}\tau_8$& $-\lambda_2$/$\lambda_5$/$-\lambda_7$  \\

\hline

\end{tabular}
\caption{Vertex factors in spinor, flavor and color spaces for the mesons and diquarks considered in this work.}
\label{vertices}
\end{table}

After performing the traces, we obtain the respective polarization functions for pseudoscalar mesons, scalar diquarks and axial diquarks,

\begin{multline}
    \Pi_{PS}^{f_1f_2}(p) = 4iN_c\Big[I_1(m_{f_1},\Lambda)+I_1(m_{f_2},\Lambda) \\+ ((m_{f_1}-m_{f_2})^2-p^2)I_2(p,m_{f_1},m_{f_2},\Lambda)\Big] \ ,
\end{multline}

\begin{multline}
    \Pi_{d,S}^{f_1f_2}(p) = 8i\Big[I_1(m_{f_1},\Lambda)+I_1(m_{f_2},\Lambda) \\+ ((m_{f_1}+m_{f_2})^2-p^2)I_2(p,m_{f_1},m_{f_2},\Lambda)\Big] \ ,
\end{multline}

\begin{multline}
    \Pi_{d,A}^{f_1f_2}(p) = -\frac{16}{3}i\Big[I_1(m_{f_1},\Lambda)+I_1(m_{f_2},\Lambda) \\+ ((m_{f_1}+m_{f_2})^2-p^2)I_2(p,m_{f_1},m_{f_2},\Lambda)\Big] \ ,
\end{multline}

where $f_1$ and $f_2$ are the two flavor of the quarks or antiquarks involved in the bound state.
Note that the diquark polarization functions do not get an additional factor $N_c$ since they do not belong to a singlet representation of ${\bm{3}_c\otimes\bm{3}_c}$. The two-line loop integral $I_2$ is expressed by,
\begin{equation}
    I_2(p,m_{f_1},m_{f_2},\Lambda) = \int^\Lambda\frac{d^4k}{(2\pi)^4}\frac{1}{[(p+k)^2-m_{f_1}^2](k^2-m_{f_2}^2) } \ .
\end{equation}
In the noncovariant cutoff scheme, and taking the bound state at rest ($\bm{p}=0$), one can reduce this expression to:
\begin{multline}
    I_2(p_0,m_{f_1},m_{f_2},\Lambda)=\\-\frac{i}{8\pi^2p_0}\bigg[m_{f_1}\mathcal{I}\Big(\sqrt{1+\frac{\Lambda^2}{m_{f_1}^2}},-\frac{p_0^2+m_{f_1}^2-m_{f_2}^2}{2p_0m_{f_1}}\Big)\\
    -m_{f_2}\mathcal{I}\Big(\sqrt{1+\frac{\Lambda^2}{m_{f_2}^2}},\frac{p_0^2+m_{f_2}^2-m_{f_1}^2}{2p_0m_{f_2}}\Big)\bigg] \ ,
\end{multline}
where the remaining integral $\mathcal{I}$ appearing in the previous expression can be given analytically,
\begin{widetext}
\begin{multline}
\mathcal{I}(M,e_0)=\int_1^{M}de\frac{\sqrt{e^2-1}}{e-e_0}
=\Theta(M-1)\Bigg(\sqrt{M^2-1} + e_0\log(M+\sqrt{M^2-1})\\
    +\begin{cases}
          \sqrt{e_0^2-1}\log\Big(\Big|\frac{M-e_0}{-1+Me_0+\sqrt{(M^2-1)(e_0^2-1)}}\Big|\Big) \quad &\text{if} ~ |e_0|\geq 1 \\
          -\sqrt{1-e_0^2}\arccos\Big(\frac{1-Me_0}{M-e_0}\Big)\quad &\text{if} ~ |e_0|< 1 \\
     \end{cases}
     +i\pi\sqrt{e_0^2-1} \ \Theta [(e_0-1)(M-e_0)]\Bigg) \ , \label{eq:Ifun}
\end{multline}
\end{widetext}
with $\Theta$ being the Heaviside step function.

\subsection{Pion and kaon decay constants~\label{app:fX}}

The pion and kaon decay constants are computed applying Eq.~(\ref{decayconst}) to the corresponding channels (see the first two lines of Table~\ref{vertices}). After performing the traces, one obtains,
\begin{equation}
    f_{\pi/K}(p)= -2iN_cg_{\pi/K\rightarrow ij}(m_i+m_j)I_2(p,m_{i},m_{j},\Lambda) \ ,
\end{equation}
where the $i$ and $j$ subscripts denote any two light flavors in the pion case, and one light flavor and one strange flavor for the kaon. The vacuum values of the decay constants are then computed assuming the meson at rest and on shell ($p_0=m_X, {\bf p}=0$).

\subsection{Quark-diquark loop functions for baryons} \label{appbaryon}

The evaluation of the baryon propagator in the quark-diquark approximation is made through the quark-diquark polarization function:

\begin{equation}
    \Pi_B^{f,d}(p) = i\int\frac{d^4k}{(2\pi)^4}S_f(p+k) \Delta_d(k) \ ,
\end{equation}
where $\Delta_d$ is the propagator of the diquark $d$, given by $T_d(p)$ in Eq.~(\ref{propagapprox}). Assuming the baryon at rest (${\bm{p}=0}$), we follow the decomposition suggested in Appendix F of Ref.~\cite{TorresRincon2015} and obtain,

\begin{widetext}
\begin{equation} \label{piB}
    \Pi_B^{f,d}(p_0) = -\frac{g_{d\rightarrow qq}^2}{8\pi^2}\left[ J_F^+(p_0) + J_F^-(p_0) + J_B^+(p_0) + J_B^-(p_0)\right] \ ,
\end{equation}
where the functions $J_{F/B}^\pm$ can be expressed analytically,

\begin{equation}
    J_F^\pm(p_0) = 
    \pm\frac{1}{2p_0}m_f^2\Bigg(\mathcal{I}\bigg(\sqrt{1+\frac{\Lambda^2}{m_f^2}},\pm\frac{m_d^2-m_f^2-p_0^2}{2p_0m_f}\bigg)
    \mp\gamma_0 \mathcal{I}^\prime\bigg(\sqrt{1+\frac{\Lambda^2}{m_f^2}}, \pm\frac{m_d^2-m_f^2-p_0^2}{2p_0m_f}\bigg)\Bigg) \ , 
\end{equation}
\begin{equation}
    J_B^\pm(p_0) =
    \frac{1}{2p_0}m_d\Bigg(\pm(m_f+\gamma_0 p_0) \mathcal{I}\bigg(\sqrt{1+\frac{\Lambda^2}{m_d^2}},\pm\frac{m_f^2-m_d^2-p_0^2}{2p_0m_d}\bigg)
    +\gamma_0 m_d \mathcal{I}^\prime\bigg(\sqrt{1+\frac{\Lambda^2}{m_d^2}},\pm\frac{m_f^2-m_d^2-p_0^2}{2p_0m_d}\bigg)\Bigg)  \ , \label{JBpm}
\end{equation}
where the function $\mathcal{I}$ is shown in Eq.~(\ref{eq:Ifun}), while the function $\mathcal{I}^\prime$ is given by,
\begin{multline}
 \mathcal{I}^\prime(M,e_0)=\int_1^{M}de\,e \frac{\sqrt{e^2-1}}{e-e_0}
 =\frac{1}{2}\Theta(M-1)\Bigg((2e_0^2+M)\sqrt{M^2-1} + (2e_0^2-1)\log(M+\sqrt{M^2-1})  \\
    +\begin{cases}
          2e_0\sqrt{e_0^2-1}\log\Big(\Big|\frac{M-e_0}{-1+Me_0+\sqrt{(M^2-1)(e_0^2-1)}}\Big|\Big) \quad &\text{if} ~ |e_0|\geq 1 \\
          -2e_0\sqrt{1-e_0^2}\arccos\Big(\frac{1-Me_0}{M-e_0}\Big)\quad &\text{if} ~ |e_0|< 1 \\
     \end{cases}
     +i\pi e_0\sqrt{e_0^2-1}\Theta[(e_0-1)(M-e_0)]\Bigg) \ .
\end{multline}
\end{widetext}

\subsection{Summary of the results} \label{Appsummary}

To summarize our results, we provide in Table.~\ref{Meanvalues} the first two moments of the posterior distributions with the weight W3 for the parameters and observables investigated in this work. 
The former can be interpreted as a good estimate of an optimal parametrization of the NJL model accounting for the constraints imposed by W3.
We indicate in addition the minimum and maximal values obtained within our prior exploration for each quantity. 

\begin{table}[htbp]
\centering
\begin{tabular}{|c|c|c|c|c|c|}
\hline
  & Unit & Mean & $\sigma$ & Min & Max  \\
\hline
$\Lambda$ & MeV & 568 & 8 & 550 & 625\\
\hline
$G\Lambda^2$ &  & 2.44 & 0.23 & 1.76 & 4.00\\
\hline
$K\Lambda^5$ &  & 10.05 & 1.35 & 4.00 & 15.0\\
\hline
$m_0$ & MeV & 5.69 & 0.42 & 4.00 & 7.00\\
\hline
$m_{0,s}$ & MeV & 134 & 5 & 95 & 150\\
\hline
$m_{u,d}$ & MeV & 515 & 53 & 315 & 1045\\
\hline
$m_s$ & MeV & 678 & 46 & 476 & 1160\\
\hline
$-\langle \overline{q}q\rangle^{\frac{1}{3}}$ & MeV & 243 & 4 & 219 & 282 \\
\hline
$-\langle \overline{s}s\rangle^{\frac{1}{3}}$ & MeV & 250 & 4 & 235 & 284\\
\hline
$m_\pi$ & MeV & 138.2 & 4.5 & 110.3 & 187.4\\
\hline
$f_\pi$ & MeV & 92.1 & 1.2 & 82.6 & 101.8\\
\hline
$m_K$ &  MeV&  497& 4 & 437 & 554\\
\hline
$f_K/f_\pi$ &  & 1.01 & 0.01 & 0.98 & 1.15\\
\hline
$m_\eta$ & MeV & 510 & 8 & 357 & 604\\
\hline
$m_{\eta^\prime}$ & MeV & 987 & 27 & 680 & 1428\\
\hline
$G_d/G$ &  & 1.53 & 0.04 & 1.00 & 1.89\\
\hline
$G_{d,V}/G$ &  & 0.97 & 0.06 & 0.75 & 1.30\\
\hline
$m_{(ud)}$ & MeV & 489 & 29 & 0 & 1442\\
\hline
$m_{(us)}$ & MeV& 682 & 22 & 368 & 1556\\
\hline
$m_{[ud]}$ & MeV & 918 & 54 & 586 & 1721\\
\hline
$m_{[us]}$  & MeV & 1060 & 49 & 728 & 1832\\
\hline
$m_{[ss]}$  & MeV & 1197 & 46 & 852 & 1944 \\
\hline
$m_N$ & MeV & 952 & 41 & 0 & 2454 \\
\hline
$m_\Lambda$ & MeV & 1099 & 40 &  292 & 2560 \\
\hline
$m_\Sigma$ & MeV & 1183 & 37 & 691 & 2577 \\
\hline
$m_\Xi$ & MeV & 1294 & 35 & 768 & 2647 \\
\hline
$m_\Delta$ & MeV & 1240 & 73 & 0 & 2376 \\
\hline
$m_{\Sigma^\star}$ & MeV & 1389 & 70 & 181 & 2500 \\
\hline
$m_{\Xi^\star}$ & MeV  & 1538 & 67 & 361 & 2624\\
\hline
$m_\Omega$&MeV  & 1688 & 64 & 539 & 2752\\
\hline
$B_{(ud)}$& MeV & 541 & 114 & 407 & 1984\\
\hline
$B_{[ud]}$& MeV  & 112 & 59 & 10 & 808 \\
\hline
$B_{N}$&MeV  & 53 & 51 & 2 & 1100 \\
\hline
$B_{\Delta}$ & MeV & 193 & 120 & 0 & 1951 \\
\hline
\end{tabular}
\caption{Summary of the posterior mean, standard deviation, and minimal and maximal values for the model parameters and predicted hadronic observables with the full weights W3.}
\label{Meanvalues}
\end{table}

\bibliography{Ma_biblio}

\end{document}